\documentclass[twocolumn, tighten, times, twocolappendix]{aastex631}

\usepackage{amsmath}

\defcitealias{Gaia_BH1}{E23}

\begin{document}

\title{ESPRESSO observations of Gaia BH1: high-precision orbital constraints and no evidence for an inner binary}

\author[0000-0002-1386-0603]{Pranav Nagarajan}
\affiliation{Department of Astronomy, California Institute of Technology, 1200 E. California Blvd., Pasadena, CA 91125, USA}

\author[0000-0002-6871-1752]{Kareem El-Badry}
\affiliation{Department of Astronomy, California Institute of Technology, 1200 E. California Blvd., Pasadena, CA 91125, USA}

\author[0000-0002-5510-8751]{Amaury H.M.J.\ Triaud}
\affiliation{School of Physics and Astronomy, University of Birmingham, Edgbaston, Birmingham B15 2TT, UK}

\author[0000-0002-3300-3449]{Thomas A.\ Baycroft}
\affiliation{School of Physics and Astronomy, University of Birmingham, Edgbaston, Birmingham B15 2TT, UK}

\author[0000-0001-9911-7388]{David Latham}
\affiliation{Center for Astrophysics $\vert$ Harvard \& Smithsonian, Harvard University, 60 Garden Street, Cambridge, MA 02138, USA}

\author[0000-0001-6637-5401]{Allyson Bieryla}
\affiliation{Center for Astrophysics $\vert$ Harvard \& Smithsonian, Harvard University, 60 Garden Street, Cambridge, MA 02138, USA}

\author[0000-0003-1605-5666]{Lars A. Buchhave}
\affiliation{DTU Space,  Technical University of Denmark, Elektrovej 328, DK-2800 Kgs. Lyngby, Denmark}

\author[0000-0003-4996-9069]{Hans-Walter Rix}
\affiliation{Max Planck Institute for Astronomy, K\"onigstuhl 17, D-69117, Heidelberg, Germany}

\author[0000-0001-9185-5044]{Eliot Quataert}
\affiliation{Department of Astrophysical Sciences, Princeton University, 4 Ivy Lane, Princeton, NJ 08544, USA}

\author[0000-0001-8638-0320]{Andrew Howard}
\affiliation{Department of Astronomy, California Institute of Technology, 1200 E. California Blvd., Pasadena, CA 91125, USA}

\author[0000-0002-0531-1073]{Howard Isaacson}
\affiliation{Department of Astronomy, University of California, Berkeley, Berkeley, CA 94720, USA}

\author[0000-0002-5945-7975]{Melissa J.\ Hobson}
\affiliation{Max Planck Institute for Astronomy, K\"onigstuhl 17, D-69117, Heidelberg, Germany}




\begin{abstract}

We present high-precision radial velocity (RV) observations of Gaia BH1, the nearest known black hole (BH). The system contains a solar-type G star orbiting a massive dark companion, which could be either a single BH or an inner BH + BH binary. A BH + BH binary is expected in some models where Gaia BH1 formed as a hierarchical triple, which are attractive because they avoid many of the difficulties associated with forming the system through isolated binary evolution. Our observations test the inner binary scenario. We have measured 115 precise RVs of the G star, including 40 from ESPRESSO with a precision of  $3$--$5$ m s$^{-1}$, and 75 from other instruments with a typical precision of $30$--$100$ m s$^{-1}$. Our observations span $2.33$ orbits of the G star and are concentrated near a periastron passage, when perturbations due to an inner binary would be largest. The RVs are well-fit by a Keplerian two-body orbit and show no convincing evidence of an inner binary. Using \texttt{REBOUND} simulations of hierarchical triples with a range of inner periods, mass ratios, eccentricities, and orientations, we show that plausible inner binaries with periods $P_{\text{inner}} \gtrsim 1.5$ days would have produced larger deviations from a Keplerian orbit than observed. Binaries with $P_{\text{inner}} \lesssim 1.5$ days are consistent with the data, but these would merge within a Hubble time and would thus imply fine-tuning. We present updated parameters of Gaia BH1's orbit. The RVs yield a spectroscopic mass function $f\left(M_{\text{BH}}\right)=3.9358 \pm 0.0002\,M_{\odot}$ --- about $7000\sigma$ above the $\sim2.5\,M_{\odot}$ maximum neutron star mass. Including the inclination constraint from {\it Gaia} astrometry, this implies a BH mass of $M_{\text{BH}} = 9.27 \pm 0.10 ~ M_{\odot}$.

\end{abstract}



\section{Introduction} \label{sec:intro}

The Milky Way is expected to contain $\sim10^8$ stellar mass black holes (BHs) \citep{brown_bethe_1994, Chawla_2022}. A tiny fraction of this population has been observed, with the presence of a BH dynamically confirmed in only $\sim$20 X-ray bright systems to date \citep{remillard_mcclintock_2006}. While BH X-ray binaries are easier to find than wider binaries hosting non-accreting (i.e.\ dormant) BHs, they are intrinsically rare, with population models estimating that only $\sim10^3$ exist in the Milky Way \citep{portegeis_zwart_1997, corral_santana_2016}. 

Searches for dormant BHs via spectroscopic techniques began even before the identification of the first BH X-ray binaries \citep{guseinov_1966, trimble_thorne_1969}. In the intervening decades, there have been extensive spectroscopic searches for dormant BHs in Galactic binaries, but only in recent years have these searches begun to yield reliable BH detections \citep{giesers_2018, shenar_2022}. Perhaps the most promising development in the search for dormant BHs thus far is the advent of precise, wide-field astrometry from the {\it Gaia} mission, which measures the positions, parallaxes, proper motions, and (potentially) binarity-induced wobble of $\sim 2$ billion stars. The mission’s 3rd data release (``DR3", \citealt{gaia_dr3_summary_2023, gaia_dr3_binaries_2023}) represents a factor of $\sim100$ increase in sample size over previous catalogs of binary orbits and is thus a promising dataset to search for various classes of rare binaries, including luminous stars orbiting dormant BH companions \citep{breivik_2017, janssens_2022}. 

Recently, \citet[][hereafter E23]{Gaia_BH1} used data from {\it Gaia} DR3 to identify Gaia BH1, a Sun-like star in a $185$-day orbit around a dark companion with inferred mass $M_2 = 9.62 \pm 0.18 ~ M_{\odot}$. At a distance of only 480 pc, the system is the nearest known BH by a factor of $\sim$3. While \citetalias{Gaia_BH1} identified the companion as a likely dormant stellar-mass BH, they noted that the data were also consistent with the unseen object being a close binary containing {\it two} BHs with total mass $\approx 9.6\,M_{\odot}$. This raised the tantalizing possibility that Gaia BH1 could contain a wider cousin of the merging BH binaries now routinely detected at cosmological distances via gravitational waves.

Indeed, a BH binary in Gaia BH1 could solve some puzzles related to the system's formation. Isolated binary evolution models struggle to form ``intermediate separation'' BH + low-mass star binaries like Gaia BH1. The current orbital separation, $a \approx 1.4$ AU, is significantly smaller than the predicted maximum radius of a solar-metallicity progenitor of a $\sim 10 ~ M_{\odot}$ BH, suggesting that the Sun-like star and BH progenitor would have interacted when the progenitor was a red supergiant. However, a Sun-like star is unlikely to successfully eject the envelope of a massive star, and is particularly unlikely to end up in an orbit as wide as 1.4\,AU if it does survive. These hurdles could potentially be avoided if the system formed as a hierarchical triple, with the solar-type star orbiting two massive stars. In that case, the two massive stars could prevent one another from expanding to red supergiant dimensions, and after two episodes of mass transfer, the Sun-like star could find itself orbiting two BHs without ever having interacted with either one. This and related scenarios have been discussed as a possible formation channel for Gaia BH1-like binaries by several works (e.g. \citetalias{Gaia_BH1}; \citealt{Chakrabarti2023, Gaia_BH2, hayashi_suto_trani_2023,dicarlo2023young}). Besides the triple scenario, other solutions for the current orbit have been proposed (see Section \ref{sec:formation_history}).

Binary population synthesis simulations predict that BH + BH binaries should be abundant, significantly outnumbering BH + star binaries \citep[e.g.][]{Shao2021}. Whether many of these binary black holes (BBHs) form with distant tertiaries is uncertain, depending on modeling of complex processes in triple evolution \citep[e.g.][]{Silsbee2017, Toonen2020}. There is little doubt, however, that many massive stars form in triples \citep{Moe2017}, and so it is natural to search for the BH binary + normal star triples they may evolve to. 

In this work, we explore and test the possibility that Gaia BH1 contains an inner BH binary. If an inner binary exists, its orbital motion would introduce non-Keplerian perturbations to the radial velocity (RV) curve of the outer star, which could be detectable with high-precision RV measurements \citep[e.g.][]{hayashi_wang_suto_2020, hayashi_suto_2020, Liu2022, hayashi_suto_trani_2023}. While the expected perturbations are small, Gaia BH1 contains a relatively bright, Sun-like star, for which RVs can indeed be measured with very small uncertainties.

The remainder of this paper is organized as follows. In Section \ref{sec:theory}, we explore the orbital parameter space of possible inner BH binaries via simulations and make predictions for the observed amplitude of the RV residuals of the outer star with respect to the Keplerian case. In Section \ref{sec:data}, we present our precision RVs for the outer star collected using ESPRESSO on the Very Large Telescope, along with a larger set of RVs for the outer star collected using various other lower resolution spectrographs. In Section \ref{sec:analysis}, we provide an updated Keplerian fit to the RV curve of the outer star and compare our derived parameters to that of \citetalias{Gaia_BH1}. While we find no convincing evidence of an inner BH binary, we also fit a hierarchical triple model to our spectroscopic data assuming the presence of a inner BBH, and derive the corresponding best-fit orbital parameters. In Section \ref{sec:discussion}, we discuss the implications of our results for the potential formation channels of Gaia BH1. Finally, in Section \ref{sec:conclusion}, we summarize our main findings and consider avenues for follow-up study on the binarity of Gaia BH1. 

\section{Expected RV signatures of a BH binary} \label{sec:theory}

We begin by summarizing the expected RV perturbations due to a BH binary for a variety of inner binary periods, mass ratios, eccentricities, and orbital configurations. 

\subsection{Circular and coplanar orbits}
\label{sec:analytic}

In the case of a coplanar and circular hierarchical triple, the RV of an outer star ($m_*$) orbiting an inner binary ($m_1$ and $m_2$) observed by a distant observer has an approximate analytic solution given by perturbation theory \citep{morais_correia_2008}. While the mean motion of the outer star about the center of mass of the inner binary is approximately Keplerian, the short-term RV modulations are non-Keplerian, with characteristic semi-amplitude given by:

\begin{equation}
    \label{eq:K_coplanar}
    K_{\text{short}} = \frac{m_1 m_2}{m_{12}^2} \left(\frac{m_{12}}{m_{123}}\right)^{5/3} \left(\frac{P_{\text{inner}}}{P_{\text{outer}}}\right)^{7/3} \left(\frac{2 \pi G m_{123}}{P_{\text{outer}}}\right)^{1/3}
\end{equation}

where $m_{12} \equiv m_1 + m_2$, $m_{123} \equiv m_{12} + m_*$, and $P_{\text{inner}}$ and $P_{\text{outer}}$ are the orbital periods of the inner and outer orbits, respectively \citep{morais_correia_2008, hayashi_wang_suto_2020}. These short-term variations are smaller than the unperturbed Keplerian semi-amplitude by a factor $\left(P_{\text{inner}} / P_{\text{outer}}\right)^{7/3}$ and have a dominant period $\approx P_{\text{inner}} / 2$ \citep{morais_correia_2008, hayashi_wang_suto_2020}. For an equal-mass ratio inner binary with $P_{\text{inner}} = 6$ days, the expected semi-amplitude for Gaia BH1 in the analytical case is $\approx 6$ m s$^{-1}$.

\subsection{Eccentric and inclined orbits}

While the prescription above provides an analytic approximation to the RV perturbations of the outer star in the simplest possible scenario, physical binaries are rarely circular or coplanar. In the case of Gaia BH1, we {\it know} that the orbit of the G star is not circular (it has eccentricity $e_{\text{outer}}\approx 0.45$; see  \citetalias{Gaia_BH1}). Moreover, we expect the inner orbit (if it exists) to be eccentric and misaligned with the outer orbit because both BHs' natal kicks would have perturbed it. In general, we expect the orbit of the outer star to precess, increasing the complexity of the dynamics of the system. While \citet{morais_correia_2011} provide analytic approximations for the cases in which the orbits of the inner binary and outer star are circular and non-coplanar or eccentric and coplanar, detailed study of the general case requires numerical integrations \citep[e.g.][]{hayashi_suto_2020}.

We calculate the expected RV perturbations in the non-circular, non-coplanar regime using \texttt{REBOUND}, a flexible N-body integrator \citep{rebound_2012}.  To begin, we use the default IAS15 integrator and uniformly sample 1000 epochs over one orbital period of the outer star (later, we will use the observing cadence of our measured RVs). In general, for each simulation, we fix the orbital parameters and mass of the outer star to the values determined by \citetalias{Gaia_BH1} (see Table \ref{tab:gaia_bh1_params}), and leave the orbital parameters and mass ratio of the inner black hole binary as free parameters. We adopt default values for any \texttt{REBOUND} parameters that are not explicitly specified here.

\begin{deluxetable*}{ccccc}
\tablecaption{Physical parameters and 1$\sigma$ uncertainties derived for the orbit of the outer Sun-like star in Gaia BH1 assuming a two-body Keplerian model. We present both the values measured by \citetalias{Gaia_BH1} and the updated values inferred in this work (see Section \ref{sec:analysis}). \label{tab:gaia_bh1_params}}
\tablehead{\colhead{Parameter} & \colhead{Description} & \colhead{Constraint (\citetalias{Gaia_BH1})} & \colhead{Median Constraint (this work)} & \colhead{MAP Solution (this work)} \\
\colhead{(1)} & \colhead{(2)} & \colhead{(3)} & \colhead{(4)} & \colhead{(5)}}
\startdata
$P$ & Orbital Period & $185.59 \pm 0.05$ days & $185.387 \pm 0.003$ days & $185.387$ \text{days} \\
$e$ & Eccentricity & $0.451 \pm 0.005$ & $0.43230 \pm 0.00002$ & $0.43230$ \\
$i$ & Inclination & $(126.6 \pm 0.4)^{\circ}$ & $(126.8 \pm 0.2)^{\circ}$ & $126.832^{\circ}$ \\
$\Omega$ & Longitude of Ascending Node & $(97.8 \pm 1.0)^{\circ}$ & $(97.0 \pm 0.7)^{\circ}$ & $96.9^{\circ}$ \\
$\omega$ & Argument of Periastron & $(12.8 \pm 1.1)^{\circ}$ & $(16.509 \pm 0.003)^{\circ}$ & $16.510^{\circ}$ \\
$T_{p}$ & Periastron Time (JD - 2457389) & $-1.1 \pm 0.7$ & $2.07 \pm 0.04$ & $2.07 $\\
$M_{\text{BH}}$ & Black Hole Mass & $9.62 \pm 0.18 ~ M_{\odot}$ & $9.27 \pm 0.10 ~ M_{\odot}$ & $9.294 ~ M_{\odot}$ \\
$M_{\text{star}}$ & Luminous Star Mass & $0.93 \pm 0.05 ~ M_{\odot}$ & $0.93 \pm 0.05 ~ M_{\odot}$ & $0.933 ~ M_{\odot} $\\
$\gamma$ & Center-of-Mass Radial Velocity & $46.6 \pm 0.6$ km s$^{-1}$ & $48.379 \pm 0.001$ km s$^{-1}$ & $48.379$ km s$^{-1}$ \\
$\beta_H$ & HIRES Radial Velocity Offset & N/A & $0.09 \pm 0.03$ km s$^{-1}$ & $0.074$ km s$^{-1}$ \\
$\beta_F$ & FEROS Radial Velocity Offset & N/A & $0.23 \pm 0.01$ km s$^{-1}$ & $0.227$ km s$^{-1}$ \\
$\beta_T$ & TRES Radial Velocity Offset & N/A & $0.51 \pm 0.01$ km s$^{-1}$ & $0.516$ km s$^{-1}$ \\
\enddata
\end{deluxetable*}

\subsubsection{Computing the RV residual curve}

In most cases, the RV curve of the outer star is indistinguishable by eye from a simple Keplerian orbit, because the perturbations are small compared to the total RV amplitude of the outer orbit. To better visualize the non-Keplerian perturbations induced by the inner binary, we fit the predicted RV curve with \texttt{RadVel} \citep{fulton_petigura_blunt_sinukoff_2018}, a code for fitting Keplerian orbits, and plot the residuals of this fit. We initialize the parameters in the $(P, T_{p}, e, \omega, K)$ basis (see Table \ref{tab:gaia_bh1_params} for descriptions of orbital parameters). We use a generic Keplerian RV model and Gaussian RV likelihood, and fit in the alternative $(P, T_c, \sqrt{e} \cos{\omega}, \sqrt{e} \sin{\omega}, K)$ basis, which imposes flat priors on all orbital elements and helps speed MCMC convergence \citep{fulton_petigura_blunt_sinukoff_2018}. Here, $T_c$ refers to the epoch of conjunction. Finally, we use the Powell minimization method with a tolerance of $10^{-5}$ to derive the best-fit orbital parameters. 

\begin{figure*}
\plotone{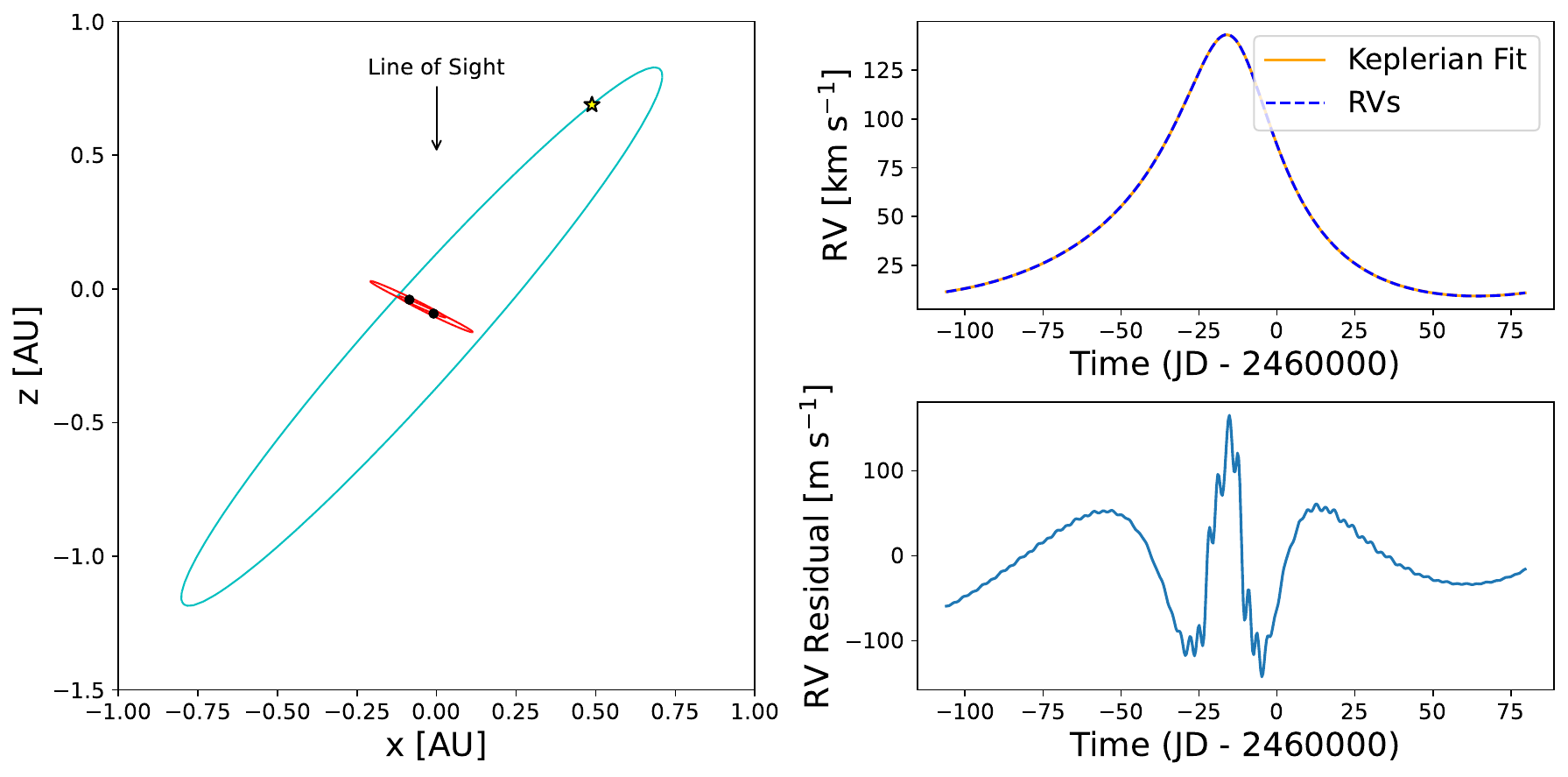}
\caption{Orbital configuration (left) and predicted RVs (right) of the hierarchical triple scenario we seek to test. RVs are measured for a Sun-like star orbiting an inner BH + BH binary. In the general case, the inner and outer orbits are both eccentric and non-coplanar. The parameters of the outer orbit are fixed to those inferred for Gaia BH1 by \citetalias{Gaia_BH1}. Here we assume an equal-mass inner binary with period $P_{\text{inner}} = 6$\,days and eccentricity $e_{\text{inner}} = 0.1$. While the RVs of the Sun-like star are nearly consistent with a Keplerian orbit (upper right), subtraction of the best-fit Keplerian orbit reveals significant residuals (lower right).  At $P_{\text{inner}} = 6$ days, the amplitude of the RV residuals is $\sim$250 m s$^{-1}$ near periastron.}
\label{fig:intro_fig}
\end{figure*}

A visualization of the RV residuals that result from this process for a typical case, where the orbital parameters of the outer star are fixed to those determined for Gaia BH1 by \citetalias{Gaia_BH1} (see Table \ref{tab:gaia_bh1_params}) and the inner equal-mass BH binary is eccentric and non-coplanar, is displayed in Figure \ref{fig:intro_fig}. We set the period, eccentricity, and inclination (relative to the line of sight) of the inner orbit to be 6 days, 0.1, and 30$^{\circ}$ respectively, with the remainder of the orbital elements matching those of the outer orbit. The qualitative behavior of the residuals is insensitive to the phase an orientation of the inner orbit. The maximum amplitude of the RV residuals, which occurs at periastron, is about 250 m s$^{-1}$. The characteristic short-timescale variability is more prominent in some portions of the orbit than others because the orbital plane of the outer star is not aligned with the orbital plane of the inner binary; consequently, the gravitational influence of the inner BBH on the outer star is stronger when the three bodies are aligned than when they are not. We also observe long-term (i.e.\ on timescales longer than $P_{\text{inner}} / 2$) variation in the RV residual curve, which we attribute to precession in the orbital parameters of the outer star.

\subsubsection{Effects of RV uncertainties and finite observing cadence}

Measuring the RV curve with the high cadence and precision shown in Figure \ref{fig:intro_fig} is currently infeasible. Here, we consider an observing strategy that is roughly representative of the observations we actually carry out for Gaia BH1 (see Section \ref{sec:data}). To initialize the \texttt{REBOUND} simulation, we fix the orbital parameters of the outer orbit to those determined by \citetalias{Gaia_BH1} and assume the same orbital parameters for the inner binary as in Figure \ref{fig:intro_fig}. Then, adopting a per-epoch RV uncertainty of 2.5 m s$^{-1}$ (as might be expected for ESPRESSO observations), we sample the RV curve of the outer star at $\sim$50 epochs spaced every few days, with increased cadence (i.e.\ every day) near periastron.

We show the resulting RV residual curve for $P_{\text{inner}} = 6$ days in Figure \ref{fig:realistic_fig}. At this inner binary period, the residuals are highly significant, with a maximum semi-amplitude of about $150$ m s$^{-1}$. This is almost two orders of magnitude larger than the assumed uncertainties, leading to a very poor fit with reduced $\chi^2 \approx 672$. In the right panel, we show the reduced $\chi^2$ value of the \texttt{RadVel} fit for inner binary periods ranging from $0.5$ to $10$ days. As expected, it is close to $1.0$ for $P_{\text{inner}} \ll 1$\,d, but rises steeply at $P_{\text{inner}} > 1$ d, when the perturbations due to the inner binary begin to exceed the RV measurement uncertainties. This suggests that we can reasonably expect to detect any inner binary with $P_{\text{inner}} \gtrsim 1$ day.

\subsubsection{Dependence on inner binary mass ratio, eccentricity, and inclination}

In Figure \ref{fig:typical_fig}, we explore how the predicted deviations from a Keplerian orbit depend on the parameters of the inner binary. In the left panels, we show the residuals after subtracting the best-fit Keplerian orbit, always assuming $P_{\text{inner}} = 6$ days. In the right panels, we show the RV residual amplitude (defined as the difference between the maximum and minimum RV residuals over the observing baseline) as a function of $P_{\text{inner}}$, with other parameters held fixed.  We fit a power law to each residual amplitude plot and report the best-fit power law index (denoted by $\alpha$) in each panel. 

Each row shows a different orbital configuration. In the top row, we simulate the case where the orbits are coplanar and both are circular, and find that the power law index is close to the theoretical expectation of $7/3=2.33$ (see Section \ref{sec:analytic}). In the next three rows, we fix the orbital parameters of the outer star to those determined by \citetalias{Gaia_BH1}, and simulate cases in which the inner orbit is circular and coplanar with the outer orbit (2nd row), circular and perpendicular to it (3rd row), and eccentric and coplanar (4th row). In these cases, the trend is not a perfect power law, and the best-fit power law index is smaller than $7/3$. This reflects the fact that the short-timescale and long-timescale perturbations have different scalings with $P_{\text{inner}}$ \citep{hayashi_suto_trani_2023}. In these cases, numerical integrations are critical, since the behavior of the RV residuals is not easy to explain analytically. Note that the RV residuals are always larger for the realistic case than for the circular + coplanar case on which most previous work has focused \citep{morais_correia_2008, hayashi_wang_suto_2020}. This is primarily because the eccentric outer orbit allows the inner BBH to get closer to the outer star at periastron. 

\subsubsection{Distinguishing an inner binary from a planet}

\begin{figure*}
\plotone{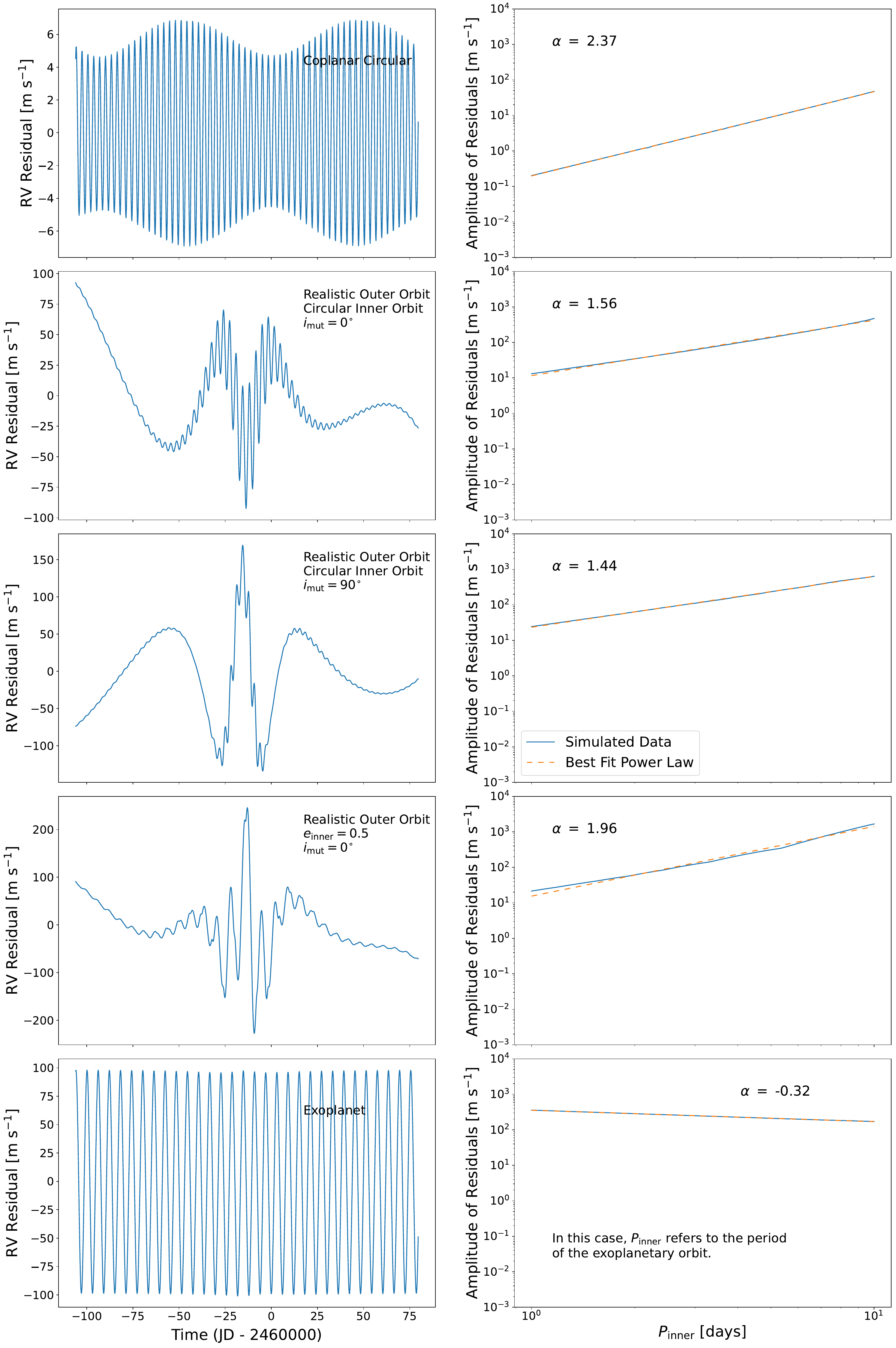}
\caption{Left: RV residuals for a variety of configurations of the hierarchical triple at $P_{\text{inner}} = 6$ days. Right: Power laws fitted to variation of amplitude of RV residuals with $P_{\text{inner}}$. We recover the theoretical expectation of a power law slope of $2.33$ in the case where the orbits are coplanar and both are circular (Equation \ref{eq:K_coplanar}). In the other cases, the power law slope is lower, but the amplitude of the RV residuals is higher at all detectable values of $P_{\text{inner}}$. Bottom panel shows a case where there is a single BH, but a Jupiter-mass planet orbits the star.}
\label{fig:typical_fig}
\end{figure*}

Finally, we simulate a case in which there is no inner BH binary, but there is a $10^{-3} ~ M_{\odot}$ planet orbiting the outer star (5th row of Figure \ref{fig:typical_fig}). Jettisoning considerations about how such a planet may have formed and survived until now, we assume it to have a circular orbit about the outer star and match the inclination and longitude of the ascending node of its orbit to that of the star's orbit. We find that the resulting RV residual curve is sinusoidal, and that the amplitude {\it decreases} with increasing period ($\alpha = -1/3$), as expected.

This case is relevant because a planet could mimic a signal from an inner binary --- indeed, perturbations due to an inner binary were considered as a false-positive for exoplanet searches before it became clear that exoplanets are more common than triple star systems with the relevant configurations \citep[e.g.][]{morais_correia_2008}. One way to distinguish between the two cases is that the amplitude of the RV variation in the exoplanetary case does not significantly increase at the periastron of the outer orbit. 

\begin{figure*}
\plotone{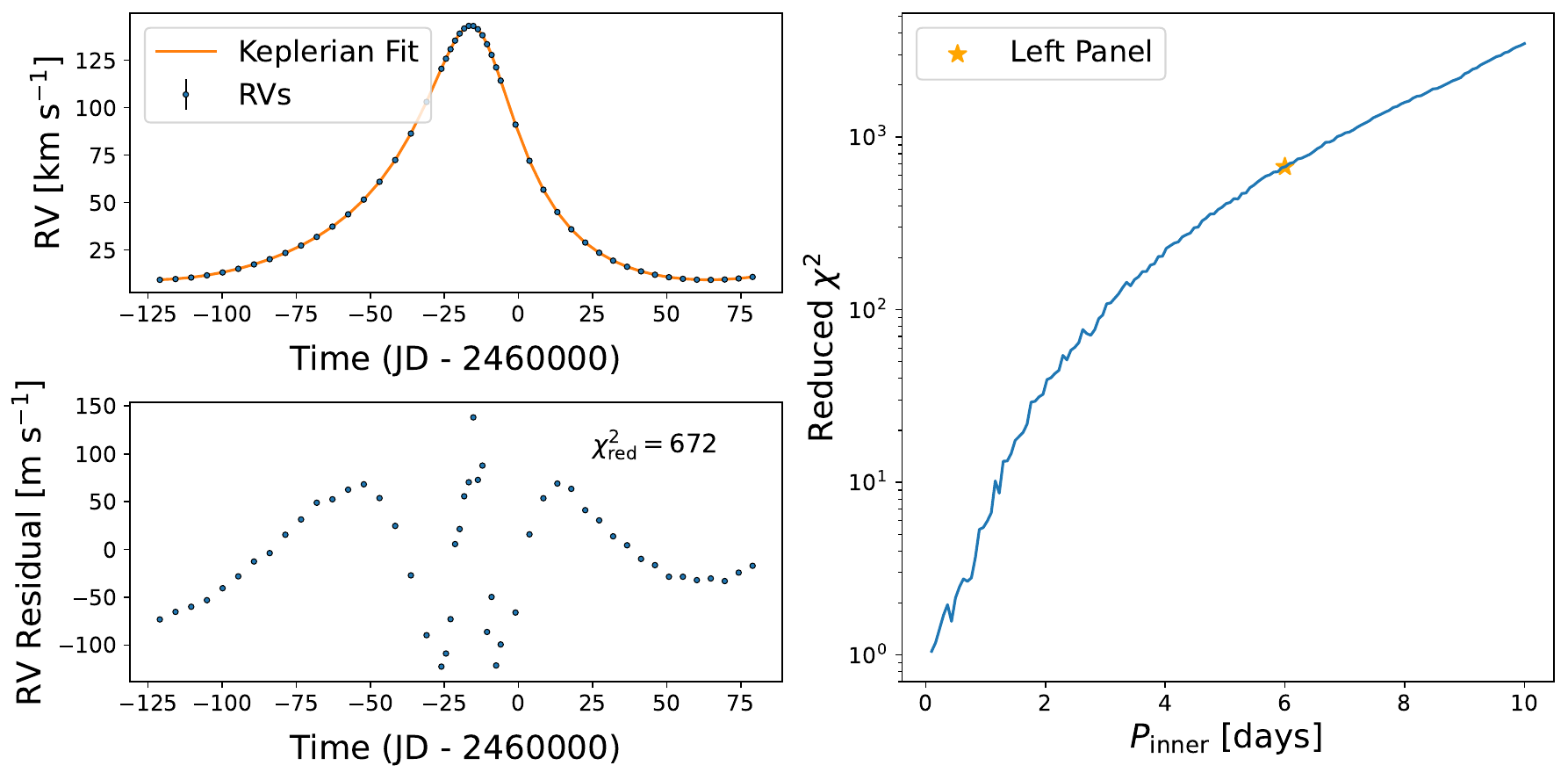}
\caption{Left: RV residuals for outer star (bottom panel) after fitting a \texttt{RadVel} Keplerian orbit to 50 noisy RVs sampled from a \texttt{REBOUND} simulation using realistic cadence (top panel). The orbital parameters of the outer star are fixed to those derived for Gaia BH1 by \citetalias{Gaia_BH1}, and the orbital parameters of the inner BBH are the same as those used for Figure \ref{fig:intro_fig}. Right: Reduced $\chi^2$ of best-fit Keplerian model as a function of period of inner binary. The reduced $\chi^2$ is close to unity at $P_{\text{inner}} \lesssim 1$ day, indicating a good fit. It rises rapidly at longer inner periods, indicating a detectable RV perturbation.}
\label{fig:realistic_fig}
\end{figure*}

\subsubsection{Short- and long-timescale RV perturbations}

\begin{figure*}
\plotone{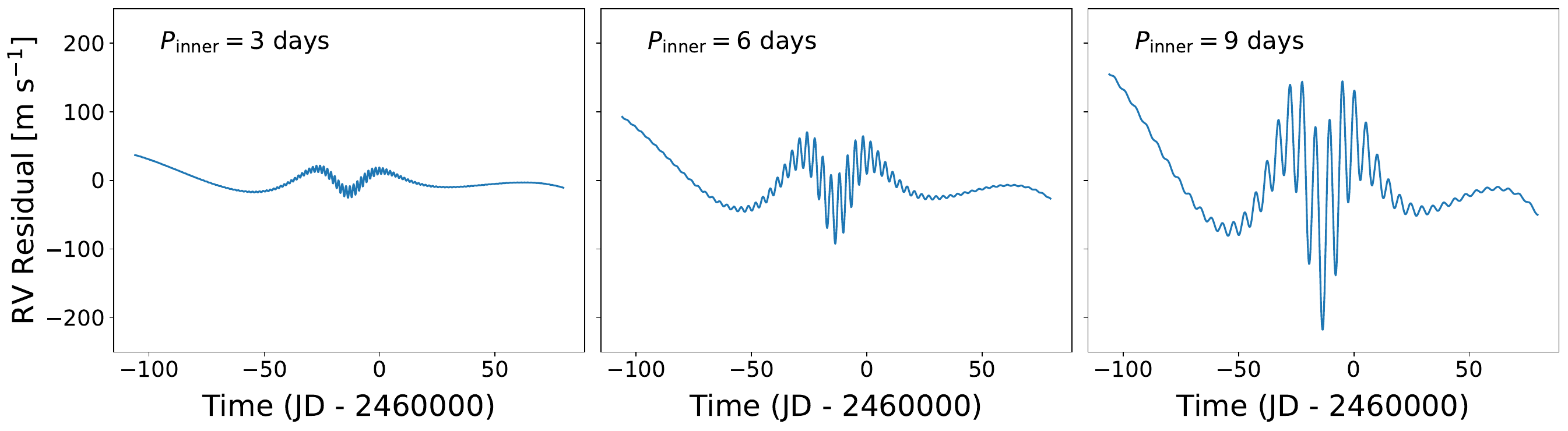}
\caption{Variation of RV residual curve shapes with $P_{\text{inner}}$. Here we fix the orbital parameters of the outer star to those determined by \citetalias{Gaia_BH1} and make the inner orbit coplanar and circular. At short $P_{\text{inner}}$, the residuals are dominated by a long-term undulation due to precession of the outer orbit. A lower-amplitude ``wobble'' on half the period of the inner binary is also present. Both the long- and short-timescale residuals become larger with increasing $P_{\text{inner}}$, but the short-timescale residuals grow more quickly, such that short- and long-timescale residuals have comparable amplitude at $P_{\text{inner}} = 9$ days.}
\label{fig:period_fig}
\end{figure*}

As is evident from the left panels of Figure \ref{fig:typical_fig}, the predicted residuals due to an inner binary show features on both short and long timescales. The short-timescale features with dominant period $P_{\text{inner}}/2$ reflect the dipole-like oscillations in the gravitational field felt by the star due to the inner binary's motion. The long-timescale residuals are a result of precession. Figure \ref{fig:period_fig} shows example residuals for a range of inner binary periods. We fix the parameters of the outer orbit to those determined by \citetalias{Gaia_BH1} and constrain the inner orbit to be coplanar and circular. The amplitude of both short- and long-timescale RV residuals increases with the orbit of the inner BBH, varying from about $50$ m s$^{-1}$ at $P_{\text{inner}} = 3$ days to $350$ m s$^{-1}$ at $P_{\text{inner}} = 9$ days. At small $P_{\text{inner}}$, the long-term perturbation due to precession is larger than the short-timescale variations. At longer $P_{\text{inner}}$, the two amplitudes are similar.

\subsubsection{Dependence on the observing time baseline}
\label{sec:baseline}

\begin{figure*}
\epsscale{0.9}
\plotone{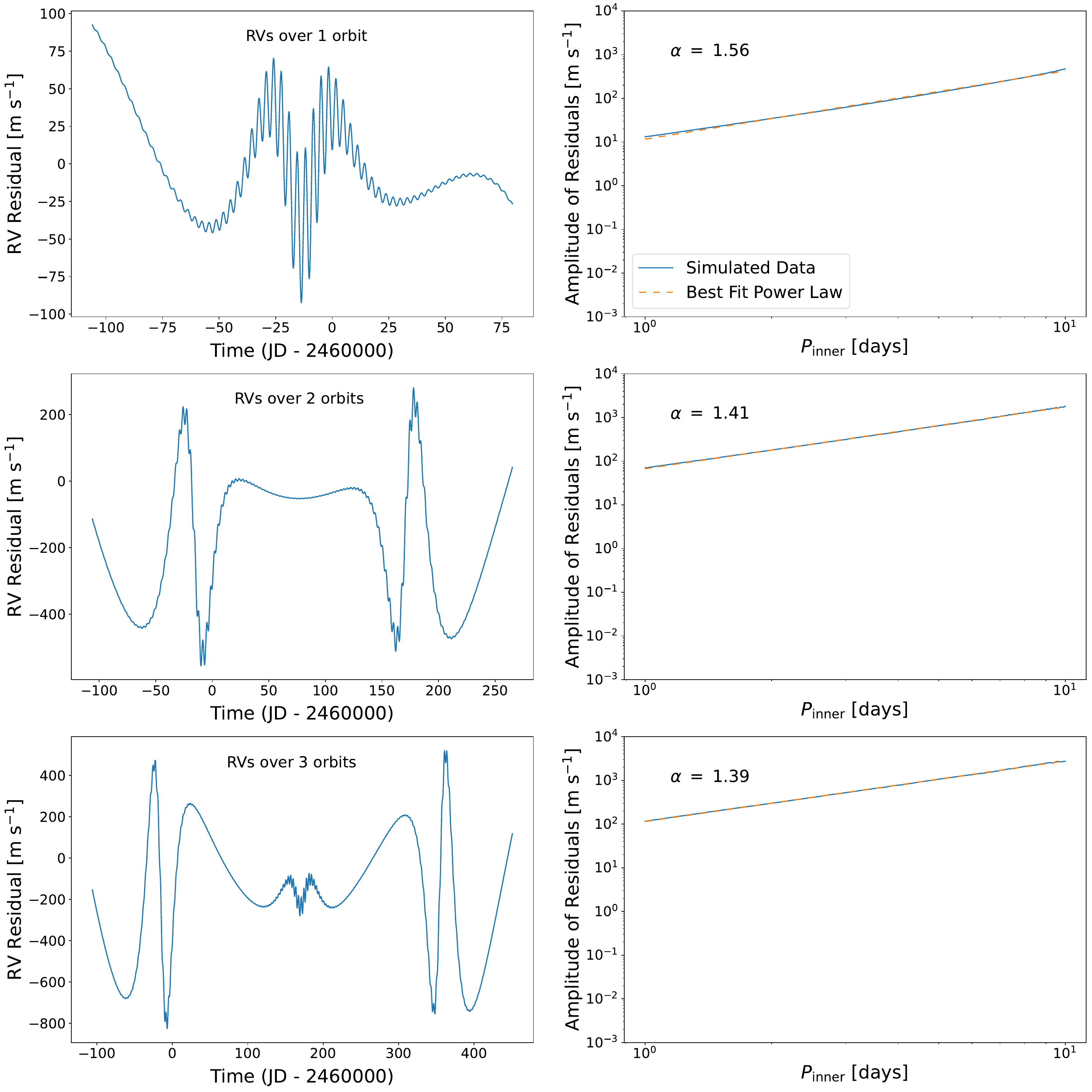}
\caption{Left: RV residuals at $P_{\text{inner}} = 6$ days for observational baselines of increasing duration in the case where we fix the orbital parameters of the outer star to those determined by \citetalias{Gaia_BH1} and constrain the inner orbit to be coplanar and circular. Right: Power laws fitted to variation of amplitude of RV residuals with $P_{\text{inner}}$ for each observational baseline. Covering more than one orbital cycle of the outer star can increase the observed RV residual amplitude significantly, because the cumulative effects of precession grow with the observing baseline.}
\label{fig:cycle_fig}
\end{figure*}

We also investigate how the amplitude of the RV residuals varies with the length of the observational baseline. In Figure \ref{fig:cycle_fig}, we again fix the orbital parameters of the outer star to those determined by \citetalias{Gaia_BH1} and constrain the inner orbit to be coplanar and circular. We then simulate observations over one, two, and three orbits. While the power law slope depends only weakly on the length of the observational baseline, extending RV coverage to two or three orbital cycles leads to significantly larger-amplitude RV residuals than coverage of one orbit. This is because precession of the outer orbit has an increasing cumulative effect when the observations cover multiple orbital cycles. We thus expect that tighter constraints on an inner binary can be obtained with a longer observing baseline.

\subsubsection{RV residual amplitudes for plausible inner binary parameters}

\begin{figure*}
\epsscale{1.05}
\plotone{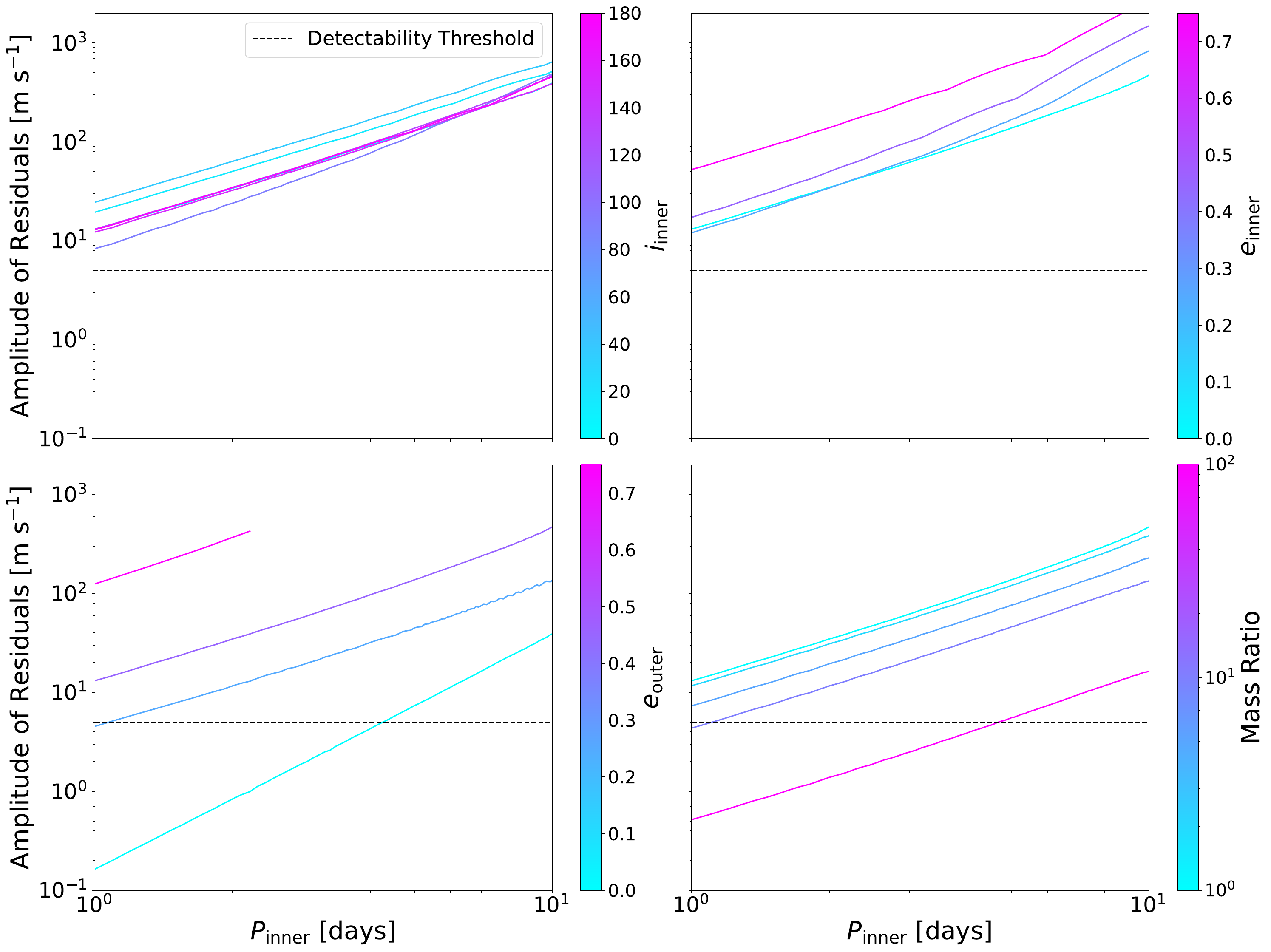}
\caption{Evolution of amplitude of RV residuals as a function of $P_{\text{inner}}$ with various orbital parameters of the inner BBH. We use the dynamical stability criterion of \citet{aarseth_mardling_2001} to restrict the power law curves from entering unstable regions in parameter space. In general, increasing the eccentricity of either orbit increases the amplitude of the observed residuals, while increasing the BBH mass ratio decreases the amplitude of the observed residuals. In addition, the residual amplitude varies non-monotonically with the orbital inclination of the inner BBH. The dashed horizontal line marks a residual amplitude of $5$ m s$^{-1}$, roughly the sensitivity of our observations. This suggests that our observations are sensitive to essentially all plausible inner binaries with $P_{\text{inner}}$ significantly above 1 day, except in the (physically dubious) case of an inner BBH mass ratio of $100$.}
\label{fig:all_power_laws}
\end{figure*}

Figure \ref{fig:all_power_laws} shows the expected RV residual amplitudes as a function of orbital period for simulations in which we systematically vary the inclination, eccentricity, and mass ratio of the inner BBH. To facilitate comparison with previous work that assumed a circular outer orbit, we also vary the outer orbit's eccentricity (though it is known for Gaia BH1). 

Increasing the eccentricity of either orbit increases the amplitude of the observed residuals. This is because the outer star reaches a smaller minimum separation relative to the inner BBH at periastron. On the other hand, increasing the mass ratio of the BHs decreases the amplitude of the observed residuals; after all, in the limit of $q_{\text{inner}} \rightarrow \infty$, we would recover the two-body Keplerian orbit. We find that the mutual inclination of the orbits has a small effect on the observed residuals, with the residual amplitude varying non-monotonically with $i_{\text{inner}}$. In almost all cases, the observed amplitude of short-term non-Keplerian RV modulations is above our detectability threshold of $\approx 5$ m s$^{-1}$ (see Section \ref{sec:data}) for $P_{\text{inner}} > 1$ day. The only exceptions are the models with $e_{\text{outer}} = 0$ (which is ruled out since we know $e_{\text{outer}} \sim 0.45$) or $q_{\text{inner}} = 100$ (which is astrophysically unlikely).

For sufficiently large inner orbits and high inner or outer eccentricities, the triple is expected to become unstable on a short timescale. We use the approximate dynamical stability criterion derived by \citet{aarseth_mardling_2001} to restrict the power law curves in Figure \ref{fig:all_power_laws} from entering regions where the hierarchical triple system is unstable. Specifically, this is relevant for large $P_{\text{inner}}$ and highly eccentric orbits; an example of this is the truncation of the $e = 0.75$ power law curve in the lower left panel of Figure \ref{fig:all_power_laws}.

\section{Data} \label{sec:data}

We now describe the observed RVs. In brief, we analyze 115 RVs obtained over 432 days, including 40 from ESPRESSO with a typical precision of  $3$--$5$ m s$^{-1}$, and 75 from FEROS, TRES, and HIRES with a typical precision of $30$--$100$ m s$^{-1}$. All of these RVs are listed in Table~\ref{tab:all_rvs}.

\subsection{FEROS}
\label{sec:feros}
We observed Gaia BH1 53 times with the Fiberfed Extended Range Optical Spectrograph \citep[FEROS;][]{Kaufer1999} on the 2.2m ESO/MPG telescope at La Silla Observatory (programs P109.A-9001, P110.A-9014, P111.A-9003, and P112.2650). The first several observations used $2\times 2$ binning to reduce readout noise at the expense of spectral resolution; the remainder used $1\times 1$ binning. The resulting spectra have resolution $R\approx 40,000$ ($2\times 2$ binning) and $R\approx 50,000$ ($1\times 1$ binning) over a spectral range of $350$--$920$ nm. Most of our observations used 1800s exposures. The typical signal-to-noise ratio (SNR) per pixel at 5800\,\AA\, is $15$.

We reduced the data using the CERES pipeline \citep{Brahm2017}, which performs bias-subtraction, flat fielding, wavelength calibration, and optimal extraction. The pipeline measures and corrects for small shifts in the wavelength solution during the course of a night via simultaneous observations of a ThAr lamp with a second fiber. We first calculate RVs by cross-correlating a synthetic template spectrum with each order individually and then report the mean RV across 15 orders with wavelengths between $4500$ and $6700$\,\AA. We calculate  the uncertainty on this mean RV from the dispersion between orders; i.e., $\sigma_{\text{RV}}\approx{\rm std}\left({\text{RVs}}\right)/\sqrt{15}$. We used a Kurucz model spectrum \citep{Kurucz1979, Kurucz1993} with $T_{\text{eff}}=5750$ K, $\log g = 4.5$, and $\rm [Fe/H]=-0.25$ from the \texttt{BOSZ} library \citep{Bohlin2017} as a template.

RVs from the first 17 observations were already presented by \citetalias{Gaia_BH1}. However, we re-reduced those data with the CERES pipeline and achieved a significantly more stable wavelength solution compared to the ESO MIDAS reduction described by \citetalias{Gaia_BH1}. The FEROS RVs analyzed here and listed in Table \ref{tab:all_rvs} thus supersede those measured by \citetalias{Gaia_BH1}. The median uncertainty of the FEROS RVs is $\approx 70$ m s$^{-1}$.

\subsection{TRES}
\label{sec:TRES}

We obtained 13 spectra using the Tillinghast Reflector Echelle Spectrograph \citep[TRES;][]{Furesz2008} mounted on the 1.5 m Tillinghast Reflector telescope at the Fred Lawrence Whipple Observatory (FLWO) atop Mount Hopkins, Arizona. TRES is a fibrefed echelle spectrograph with a wavelength range of $390$--$910$ nm and spectral resolution $R\approx 44,000$. Exposure times ranged from 1800 to 3600 seconds. We extracted the spectra as described in \citet{Buchhave2010}.

As with the FEROS data, we measured RVs by cross-correlating the normalized spectra from each of 31 orders with a template, and we estimate RV uncertainties from the dispersion between RVs measured from different orders; i.e., $\sigma_{\text{RV}}={\rm std}\left(\text{RVs}\right)/\sqrt{31}$. We used the same Kurucz template from the \texttt{BOSZ} library that we used for the FEROS data ($T_{\text{eff}}=5750$ K, $\log g = 4.5$,  $\rm [Fe/H]=-0.25$). The median uncertainty of the TRES RVs is $\approx 50$ m s$^{-1}$.

\subsection{HIRES}
\label{sec:hires}

We analyze 9 telluric-calibrated RVs measured with the High Resolution Echelle Spectrometer \citep[HIRES;][]{1994SPIE.2198..362V} mounted on the 10 m Keck telescope at W.\ M.\ Keck Observatory at Mauna Kea, Hawaii. These are the same HIRES RVs that were analyzed by \citetalias{Gaia_BH1}. We adopt an uncertainty of $100$ m s$^{-1}$ for all of these RVs. 

\subsection{ESPRESSO}
\label{sec:espresso}

We observed Gaia BH1 40 times with the Echelle SPectrograph for Rocky Exoplanets and Stable Spectroscopic Observations \citep[ESPRESSO;][]{Pepe2021} at the VLT (program 111.24GP.001). We used singleHR mode, in which the instrument can observe with any of the four 8.1 m Unit Telescopes (UTs). We used $2\times 1$ binning, yielding a typical spectral resolution $R=140,000$  over a wavelength range of $380$--$686$ nm, and used 900s exposures for all observations. The typical SNR at $600$ nm was $20$ and ranged from $15$--$25$, depending on seeing and lunar phase. The ESPRESSO observations are spread over a $170$ day period, spanning most of one orbit. The typical cadence was one observation every $3$--$6$ days away from periastron, and nearly daily observations near periastron. Gaps in coverage near periastron were due to bad weather or scheduling constraints.

\begin{figure*}
\plotone{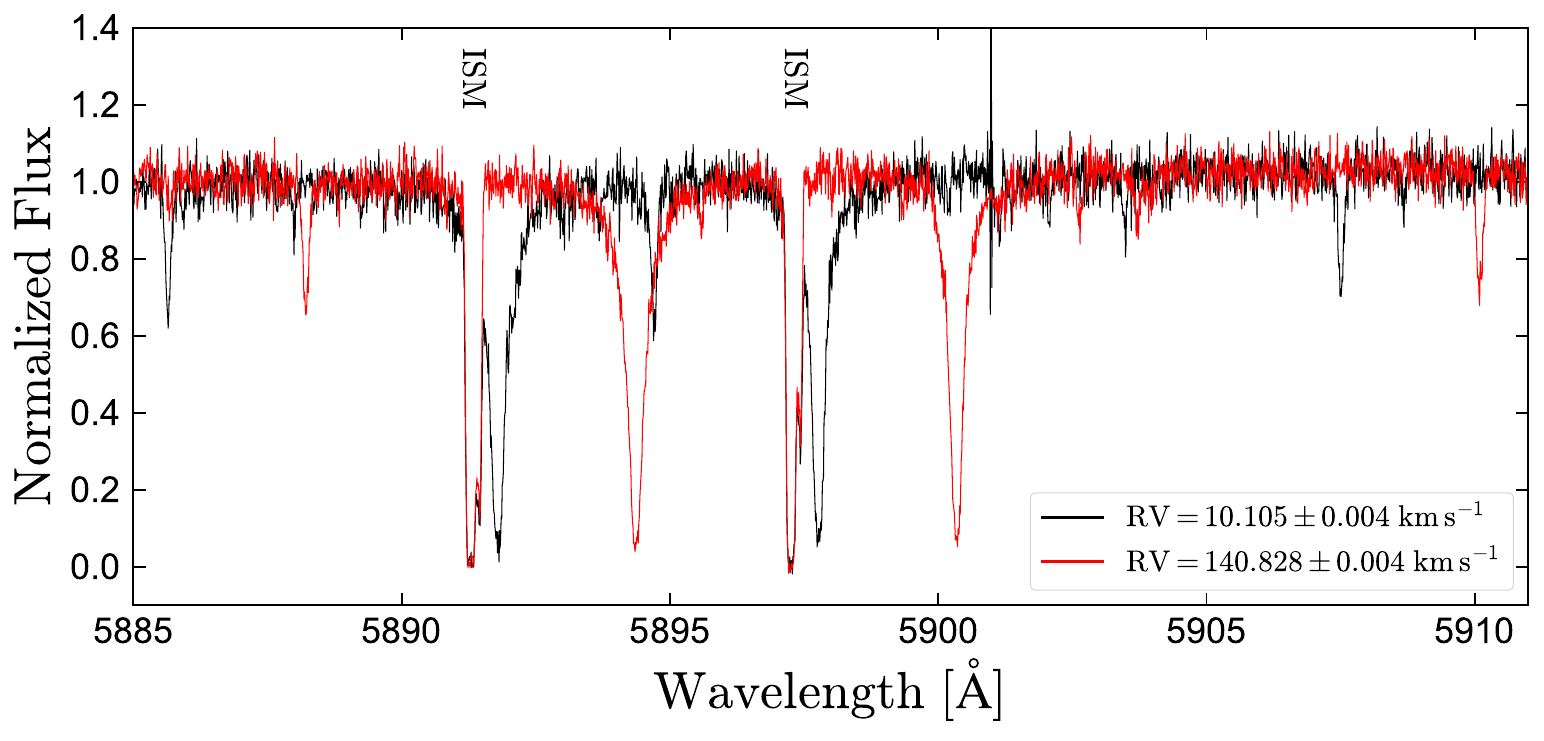}
\caption{Typical ESPRESSO spectra of Gaia BH1 from the two epochs with extremal RVs. Both spectra have SNR $\sim$30 per pixel, leading to $\sim 4$ m s$^{-1}$ RV uncertainties. This cutout is centered on the Fraunhofer Na ``D'' lines. Both photospheric and interstellar lines are present. The photospheric lines trace the RV of the Sun-like star. The interstellar lines (labeled ``ISM'') are very sharp and stable between epochs, highlighting the high resolution of the data ($R\approx 140,000$) and the stability of the wavelength solution.}
\label{fig:example_espresso_spectra}
\end{figure*}

We reduced the data using version 3.0.0 of the \texttt{ESPRESSO DRS} pipeline maintained by ESO and executed through the \texttt{EsoRex} software. We subsequently measured RVs using the \texttt{espda\_compu\_radvel} routine within the \texttt{ESPRESSO DAS} package (version 1.3.7). This routine cross-correlates all orders of the extracted spectra with a G2 weighted binary mask and then fits the resulting cross-correlation function (CCF) with a Gaussian. The mask gives maximum weight to strong lines and masks regions of the spectra affected by telluric absorption, chromospheric activity, and interstellar absorption \citep[e.g.][]{Pepe2002, Lafarga2020}. We evaluated the CCFs on a $0.5$ km s$^{-1}$ grid.  RV uncertainties are calculated from the curvature of the CCF \citep[see][their Appendix A]{Boisse2010}. The median RV uncertainty is $4.5$ m s$^{-1}$. Examples of two typical spectra from our program are shown in Figure~\ref{fig:example_espresso_spectra}. 

The high resolution of the ESPRESSO data allows us to set a more stringent limit on the projected rotation velocity of the G star than could be set previously: comparing a SNR $\sim 200$ co-added rest-frame spectrum of the G star to Kurucz model spectra broadened with the rotational kernel from \citet{Gray2008}, we estimate $v\sin i < 2$ km s$^{-1}$. The actual value of $v \sin i$ could be significantly lower than this; uncertainty in the star's micro/macroturbulent velocities limits an even more precise measurement. 

\subsection{Summary of RVs}

We simultaneously analyze lower-precision ($30$--$100$ m s$^{-1}$) RV measurements from FEROS, TRES, and HIRES, and higher-precision ($3$--$5$ m s$^{-1}$) RV measurements from ESPRESSO, with coverage over 432 days (2.33 orbits). The RVs are shown in Figure~\ref{fig:updated_astro_fig} and listed in Table~\ref{tab:all_rvs}. 

In addition to the FEROS and HIRES RVs discussed above, \citetalias{Gaia_BH1} also included several low-precision (uncertainties $\sigma_{\text{RV}} \gtrsim 1$ km s$^{-1}$) RVs from LAMOST, GMOS, X-SHOOTER, MagE, and ESI. We include these data in Figure \ref{fig:updated_astro_fig} and Table \ref{tab:all_rvs} but do not include them in the fit because they are much less precise than the other data and would require 5 additional free parameters (i.e.\ instrumental offsets). 

We also obtained 16 spectra with the Keck Planet Finder \citep[][]{Gibson2016}, a new high-precision RV spectrograph installed on the Keck telescope. These data should in principle allow RV measurement with uncertainties comparable to ESPRESSO. However, we have not yet succeeded in obtaining long-term stability in the instrument's wavelength solution due to software issues, and thus defer analysis of these data to future work. 

\section{Analysis} \label{sec:analysis}

We built a predictive model for the RV variation of the outer star in Gaia BH1 in both the two-body Keplerian and hierarchical triple cases. In the Keplerian case, we included the center-of-mass RV, component masses, orbital elements (period, eccentricity, inclination, argument of periastron, longitude of the ascending node, and periastron time) of the outer orbit, and instrumental offsets (with respect to ESPRESSO) as free parameters. In the hierarchical triple case, we used \texttt{REBOUND} simulations in place of an analytical RV model, and additionally included the orbital elements and mass ratio of the inner BBH as free parameters. 

In both cases, we placed a Gaussian prior on the mass of the Sun-like star, with a mean of $0.93 ~ M_{\odot}$ and a standard deviation of $0.05 ~ M_{\odot}$. For all other parameters, we used uniform priors. The log-likelihood assumes Gaussian uncertainties:

\begin{equation}
    \ln L_{\text{RVs}} = - \frac{1}{2} \sum_t \left(\frac{\text{RV}_{\text{obs}, t} - \text{RV}_{\text{pred}, t}}{\sigma_{\text{RV}, t}}\right)^2
\end{equation}

In order to constrain the inclination and companion mass, we jointly fit the constraints on the star's plane-of-the-sky motion from {\it Gaia}. Our approach is very similar to \citetalias{Gaia_BH1}. The \textit{Gaia} astrometric solution is given by joint constraints on 12 astrometric parameters: right ascension, declination, proper motion in right ascension, proper motion in declination, parallax, orbital period, periastron time, eccentricity, and the four Thiele-Innes elements, $A$, $B$, $F$, and $G$ \citep{halbwachs_2023}. Given the best-fit {\it Gaia} astrometric parameters $\mu_{\text{ast}}$, their covariance matrix $\Sigma_{\text{ast}}$, and the vector of predicted astrometric parameters for a given likelihood call, $\theta_{\text{ast}}$, the astrometric log-likelihood can be expressed as:

\begin{equation}
    \ln L_{\text{ast}} = -\frac{1}{2} (\theta_{\text{ast}} - \mu_{\text{ast}})^T \Sigma_{\text{ast}} (\theta_{\text{ast}} - \mu_{\text{ast}}).
\end{equation}

We calculate $\theta_{\text{ast}}$ under the assumption that the companion is dark.
The total log-likelihood function includes both the astrometric and RV terms: 

\begin{equation}
    \label{eq:lnL_tot}
    \ln L = \ln L_{\text{ast}} + \ln L_{\text{RVs}}
\end{equation}

Our joint fitting of the astrometry and RVs in the two-body Keplerian case is almost identical to \citetalias{Gaia_BH1}, with the following minor modification. \citetalias{Gaia_BH1} left the source's right ascension, declination, proper motions, and parallax as free parameters during fitting, to be constrained only by the {\it Gaia} astrometric solution. To reduce the number of free parameters, we excise these parameters from $\theta_{\text{ast}}$ and $\mu_{\text{ast}}$, and remove the corresponding rows and columns from $\Sigma_{\text{ast}}$. In predicting the Thiele-Innes elements, we assume a parallax of $\varpi=2.09$ mas. We verified that this simplification (which reduces the dimensionality of the fit) speeds up convergence while having no significant effect on the constraints on the parameters of interest. 

We began by fitting a two-body Keplerian orbit. Then, we added an additional set of orbital parameters, and tried a hierarchical triple fit instead. We finally considered whether the improvement in the log posterior probability was sufficient to warrant the extra free parameters (i.e.\ added model complexity). We describe this procedure and our results below.

\subsection{Updated fit with astrometry for Gaia BH1}

\begin{figure*}
\plotone{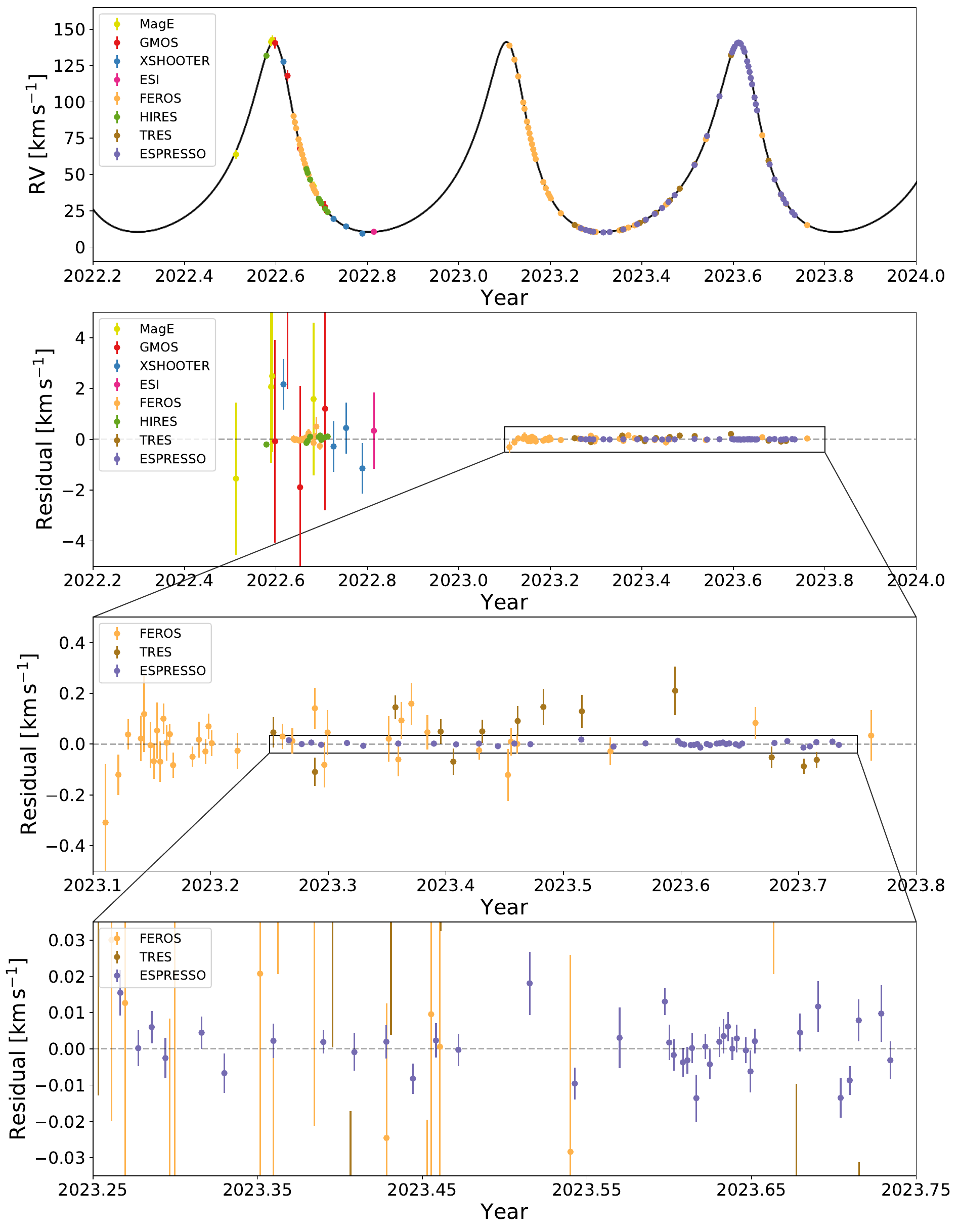}
\caption{Best-fit RV curve for Gaia BH1 based on \textit{Gaia} DR3 astrometry and our updated spectroscopic measurements, assuming a two-body Keplerian orbit. The top panel shows the observed data points over 50 RV curves randomly sampled from the posterior. The bottom panels show the residuals relative to the MAP solution plotted at three levels of precision, with the last panel focusing on the latest ESPRESSO measurements. At all levels of precision, the RVs are generally consistent with the best-fit model at the $1$--$2\sigma$ level.}
\label{fig:updated_astro_fig}
\end{figure*}

Assuming a Keplerian outer orbit and a single inner black hole, we used ensemble MCMC sampling \citep[\texttt{emcee};][]{emcee_2013} with 64 walkers and 8000 total iterations to derive updated best-fit parameters for Gaia BH1. We show the best-fit Keplerian RV curve and corresponding residuals in Figure \ref{fig:updated_astro_fig}. Compared to \citetalias{Gaia_BH1}, our data now cover three orbital cycles, and our typical RV uncertainties are a factor of 100--1000 smaller, leading to much tighter constraints. 

We show the resulting RV residuals at increasing levels of precision in the remaining three panels of Figure \ref{fig:updated_astro_fig}. At each level of precision, we see that most of the RV residuals are consistent with zero to within 1--2$\sigma$ (given the reported uncertainties in our measurements). From the third level of the residual plots, we confirm that this also holds true for the highest-precision ESPRESSO data. The small scatter in the ESPRESSO RV residuals implies that the luminous star in Gaia BH1 has low RV jitter (due to e.g.\ convection, pulsations, or other systematics) of at most $\approx 3$ m s$^{-1}$ (see \citealt{luhn_2020} for a discussion). Given the small residuals in the ESPRESSO data, it is clear that the scatter in the less-precise FEROS and TRES data acquired over the same time period is dominated by noise. However, these medium-precision RVs are still quite important for our results, because they cover three orbits and thus tightly constrain the orbital period, while the ESPRESSO data cover only one. Based on the lack of obvious structure in the residuals from a two-body orbit, there is no immediate evidence for deviations from a Keplerian orbit. 

\begin{figure*}
\plotone{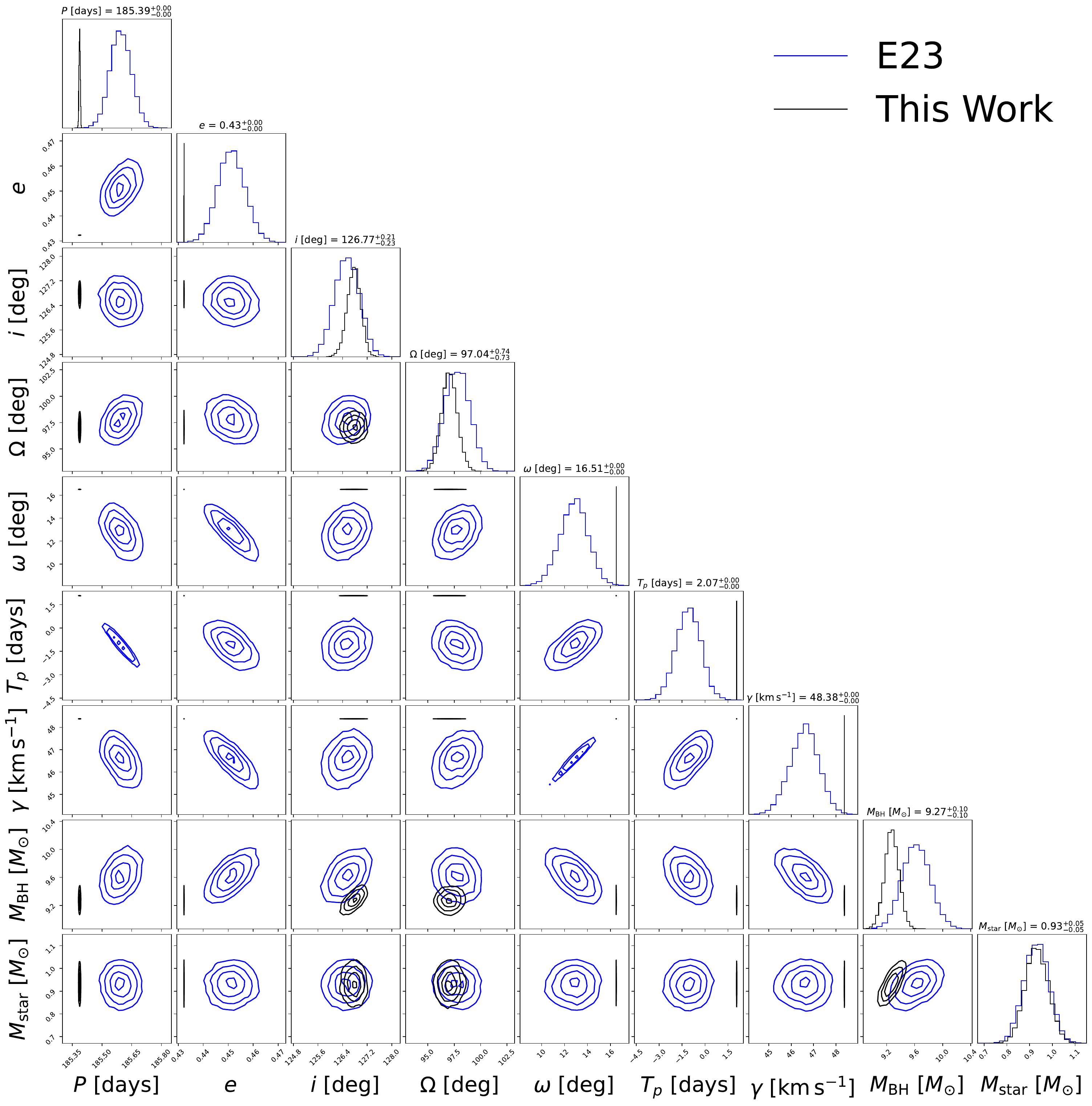}
\caption{Corner plot for orbital parameters of Gaia BH1 from MCMC sampling based on \textit{Gaia} DR3 astrometry and our updated spectroscopic measurements, assuming a two-body Keplerian orbit. The diagonal entries display the marginal distribution of each parameter, while each of the other panels displays a joint distribution. The new constraints (black) are plotted over the constraints from \citetalias{Gaia_BH1} (blue). Our new data allow us to place much tighter constraints on the orbital parameters than those obtained by \citetalias{Gaia_BH1}. The best-fit parameters are similar to those of \citetalias{Gaia_BH1} in an absolute sense, but several parameters are in tension at the $3$--$4\sigma$ level (see Table \ref{tab:gaia_bh1_params}). }
\label{fig:overlaid_corner_plot}
\end{figure*}

In Figure \ref{fig:overlaid_corner_plot}, we compare our constraints from the fit described above (black contours) to those from \citetalias{Gaia_BH1}. The  period, eccentricity, argument of periastron, periastron time, and center-of-mass RV of the orbit of the outer star are much more tightly constrained than they were by  \citetalias{Gaia_BH1}. The uncertainties in $i$, $\Omega$, and $M_{\text{BH}}$ have decreased only slightly. These parameters are constrained most directly by astrometric data, which has not changed, but they are covariant with other parameters. The constraint on $M_{\text{star}}$ comes from our SED-informed prior and is unchanged. We report our updated best-fit values in Table \ref{tab:gaia_bh1_params}. Compared to the \citetalias{Gaia_BH1} constraints,  we find that the orbital period, eccentricity, and inferred companion mass have all decreased, while the argument of periastron, periastron time, and center-of-mass RV have increased. Our new value for the companion mass is $M_{\text{BH}} = 9.27 \pm 0.10 ~ M_{\odot}$, corresponding to an RV semi-amplitude $K_{\text{star}} = 65.3785 \pm 0.0009$  km s$^{-1}$ and a spectroscopic mass function $f\left(M_{\text{BH}}\right) = 3.9358 \pm 0.0002 ~ M_{\odot}$. Our derived values for the orbital period, eccentricity, argument of periastron, periastron time, and center-of-mass RV differ from the values found by \citetalias{Gaia_BH1} by about $4.1\sigma$, $3.7\sigma$, $3.4\sigma$, $4.5\sigma$ and $3.0\sigma$ respectively. The remaining orbital parameters are consistent with each other to within 2$\sigma$. 

The small but significant tension between our best-fit orbital parameters and those of \citetalias{Gaia_BH1} suggests that the uncertainties reported by \citetalias{Gaia_BH1} were somewhat underestimated. There are a few possible reasons this could have occurred. \citetalias{Gaia_BH1} did not fit for instrumental RV offsets, and offsets between the seven spectrographs they used could be significant. However, the RVs analyzed by \citetalias{Gaia_BH1} are nearly all consistent with our updated solution at the $1$--$2\sigma$ level (2nd panel of Figure~\ref{fig:updated_astro_fig}), suggesting that RVs are unlikely to be the main source of the disagreement. Another possibility is that the {\it Gaia} astrometric uncertainties are underestimated somewhat \citep[see also][]{Chakrabarti2023}. The RVs obtained by \citetalias{Gaia_BH1} covered less than one orbital cycle, so the period constraint from their joint fit came primarily from astrometry. In contrast, our RVs cover three orbital cycles with high precision; so our constraint on the period and periastron time come almost entirely from the RVs. 

Because our measured RVs are generally consistent with the best-fit RV curve within their uncertainties (i.e., we achieve a reduced $\chi^2 = 1.69$), we are confident that the parameters we infer from RVs directly (e.g., period, eccentricity, and mass function) have robust uncertainties. The uncertainties on the parameters that depend mainly on the astrometric orbit --- inclination, longitude of the ascending node, and BH mass --- are harder to assess, and may be underestimated. It will become easier to robustly estimate uncertainties on these quantities when epoch astrometry is published in {\it Gaia} DR4. For now, the parameters reported in Table \ref{tab:gaia_bh1_params} supersede those reported by \citetalias{Gaia_BH1}, and we adopt them for the two-body solution in the rest of this work.

Note that the reported $\gamma$ in Table \ref{tab:gaia_bh1_params} represents the system's center-of-mass velocity {\it relative to our adopted ESPRESSO zeropoint}. That uncertainty in that zeropoint -- which must be accounted for in e.g.\ kinematic analysis of the system's Galactic orbit or comparison with precision RVs from other instruments or templates -- is significantly larger, on the order of $\sim 100$ m s$^{-1}$ (see \citealt{lindegren_2003} for a discussion). 

\subsection{Hierarchical triple fit for Gaia BH1}

Next, we tried fitting the data with a hierarchical triple model implemented with \texttt{REBOUND}. The model is relatively high-dimensional, with 19 free parameters (including instrumental offsets). We find the posterior to be multimodal and challenging to sample from. In the limits of $P_{\text{inner}} \to 0$ or $q_{\text{inner}} \to 0$, we recover the two-body Keplerian solution. For $0.5 \lesssim P_{\text{inner}}/\text{day} \lesssim 1.3$, there are multiple distinct combinations of inner binary parameters that yield slightly better fits (i.e., higher posterior probabilities) than the best-fit two-body solution (which, after all, has 7 fewer free parameters). For $P_{\text{inner}} \gtrsim 1.5$ days, the inner binary perturbs the outer orbit too much, leading to a poorer fit than the two-body solution. Complex posteriors are common when fitting RV data, and robust algorithms have been developed to sample from them \citep[e.g.][]{Price-Whelan2017}. Compared to the standard problem of fitting a two-body orbit, the hierarchical triple fit has the added challenges that (a) the space is higher-dimensional, with coupled inner and outer orbits, making brute-force approaches like rejection sampling infeasible, and (b) each call to the likelihood function requires running a \texttt{REBOUND} simulation, which is more expensive than evaluating the quasi-analytic equations required to predict RVs in the two-body case.

We experimented with a variety of sampling approaches, including directly running ensemble sampling with $\texttt{emcee}$, running ensemble sampling with $\texttt{emcee}$ after using an optimizer to determine a favorable initialization, and directly running dynamic nested sampling with $\texttt{dynesty}$. Ultimately, we settled on the approach described below.

We used dynamic nested sampling \citep[\texttt{dynesty};][]{dynesty_2020} to sample from the 19-dimensional posterior distribution. The free parameters of our fit are listed in Table \ref{tab:gaia_bh1_triple}. Compared to the two-body fit, the extra parameters are the period, eccentricity, inclination, longitude of the ascending node, argument of periastron, periastron time, and mass ratio of the inner orbit. As with the two-body fits, we added an astrometric term to the likelihood that compared the parameters of the outer orbit to the {\it Gaia} constraints (Equation \ref{eq:lnL_tot}). Since the {\it Gaia} solution assumes a two-body orbit, we treated the inner binary as a point mass when calculating the Thiele-Innes elements of the outer orbit. We used the random walk sampling method with $500$ live points and set a maximum of $10^6$ likelihood calls. We found that the sampler would hit the default stopping condition (defined to be when the estimated remaining evidence falls below the default threshold, see \citealt{dynesty_2020}) slightly before achieving the maximum number of likelihood calls. Consequently, we do not expect better performance with a larger value for the maximum number of likelihood calls. We assumed truncated normal priors on the parameters of the outer orbit based on the results of the two-body Keplerian fit (see Table \ref{tab:gaia_bh1_params}), and flat, broad priors on orbital parameters of the inner binary. We set a lower limit of $P_{\text{inner}} > 0.5$ days because likelihood calls became prohibitively expensive in the limit of $P_{\text{inner}} \to 0$. 

We ran several \texttt{dynesty} runs with different sampling methods, initializations, and slight variations in the priors, and then compared the maximum posterior probabilities and marginalized constraints on inner binary parameters across runs. While the maximum probabilities were relatively stable across runs (with $\ln P_{\max}$ varying by $\lesssim 1$), the best-fit parameters varied between runs by more than their formal uncertainties. This suggested that the runs were not fully converged. To explore the posteriors more thoroughly in the vicinity of the maximum probability solutions, we initialized an \texttt{emcee} chain with 64 walkers and 3125 steps at the maximum-probability sample from each \texttt{dynesty} run. These chains achieved slightly higher posterior probabilities than the best \texttt{dynesty} samples. We report the highest-probability solution achieved across all such runs as the MAP (maximum a posteriori) solution in Table \ref{tab:gaia_bh1_triple}, where we also report the marginalized median and middle 68\% constraints from the corresponding \texttt{emcee} run. The MAP solution falls within the marginalized middle 68\% range for all parameters. The best-fit solution is consistent with an inner BH binary with $P_{\text{inner}} \lesssim 1.5$ days.

We emphasize that the posterior distribution of the hierarchical triple fit is complex, and that --- because our sampling is likely not fully converged --- there is no guarantee that our reported solution corresponds to the best absolute solution, or that the reported uncertainties encompass all possible solutions. However, given the small residuals of both the two-body and three-body fits (Figure \ref{fig:models_residuals}), we consider it unlikely that a significantly better three-body solution exists. For the best-fit solution we report, the improvements over the best two-body solution are marginal.

\begin{deluxetable*}{cccc}
\tablecaption{Best-fit parameters for the orbit of the outer Sun-like star in Gaia BH1 assuming a hierarchical triple model. Error bars on the median solution are derived from the 16th and 84th percentiles. \label{tab:gaia_bh1_triple}}
\tablehead{\colhead{Parameter} & \colhead{Description} & \colhead{Median Derived Value (this work)} & \colhead{MAP Derived Value (this work)} \\
\colhead{(1)} & \colhead{(2)} & \colhead{(3)} & \colhead{(4)}}
\startdata
$P_{\text{outer}}$ & Period of Outer Orbit &    $185.45_{-0.02}^{+0.01}$ days & $185.4594$ days\\
$e_{\text{outer}} $ & Eccentricity of Outer Orbit & $0.43243_{-0.00005}^{+0.00005}$ & $0.43245$\\
$i_{\text{outer}} $ & Inclination of Outer Orbit & $(126.7_{-0.2}^{+0.2})^{\circ}$ & $126.653^{\circ}$\\
$\Omega_{\text{outer}} $ & Longitude of Ascending Node of Outer Orbit & $(98.2_{-0.9}^{+0.9})^{\circ}$ & $98.2^{\circ}$\\
$\omega_{\text{outer}} $ & Argument of Periastron of Outer Orbit &  $(16.518_{-0.007}^{+0.008})^{\circ}$ & $16.521^{\circ}$\\
$T_{p, \text{outer}} $ & Periastron Time (JD - 2460000) of Outer Orbit & $-13.506_{-0.005}^{+0.005}$ & $-13.505$ \\
$P_{\text{inner}}$ & Period of Inner Orbit & $0.9_{-0.1}^{+0.1}$ days & $0.98$ days \\
$e_{\text{inner}} $ & Eccentricity of Inner Orbit & $0.19_{-0.09}^{+0.08}$ & $0.205$\\
$i_{\text{inner}} $ & Inclination of Inner Orbit & $(108.6_{-5.0}^{+4.4})^{\circ}$ & $108.8^{\circ}$\\
$\Omega_{\text{inner}} $ & Longitude of Ascending Node of Inner Orbit & $(171.0_{-4.9}^{+3.4})^{\circ}$ & $172.4^{\circ}$\\
$\omega_{\text{inner}} $ & Argument of Periastron of Inner Orbit & $(8.6_{-6.2}^{+11.6})^{\circ}$ & $6.5^{\circ}$\\
$T_{p, \text{inner}} $ & Periastron Time (JD - 2460000) of Inner Orbit & $-0.7_{-0.2}^{+0.1}$ & $-0.70$\\
$q_{\text{inner}}$ & Mass Ratio of Black Hole Binary & $0.8_{-0.2}^{+0.1}$ & $0.80$ \\
$M_{\text{BH, tot}}$ & Total Mass of Black Hole Binary & $9.24_{-0.08}^{+0.08} ~ M_{\odot}$ & $9.235 ~ M_{\odot}$\\
$M_{\text{star}}$ & Mass of Luminous Star & $0.93_{-0.05}^{+0.05} ~ M_{\odot}$ & $0.9337 ~ M_{\odot}$\\
$\gamma$ & Center-of-Mass Radial Velocity &  $48.375_{-0.002}^{+0.002}$ km s$^{-1}$ & $48.375$ km s$^{-1}$ \\
$\beta_H$ & HIRES Radial Velocity Offset &  $0.10_{-0.04}^{+0.03}$ km s$^{-1}$ & $0.101$ km s$^{-1}$ \\
$\beta_F$ & FEROS Radial Velocity Offset & $0.23_{-0.01}^{+0.01}$ km s$^{-1}$ & $0.224$ km s$^{-1}$ \\
$\beta_T$ & TRES Radial Velocity Offset &  $0.52_{-0.01}^{+0.02}$ km s$^{-1}$ & $0.525$ km s$^{-1}$ \\
\enddata
\end{deluxetable*}

\section{Discussion} \label{sec:discussion}

\subsection{Limits on period of inner BH binary}
\label{sec:limits}

\begin{figure*}
\plotone{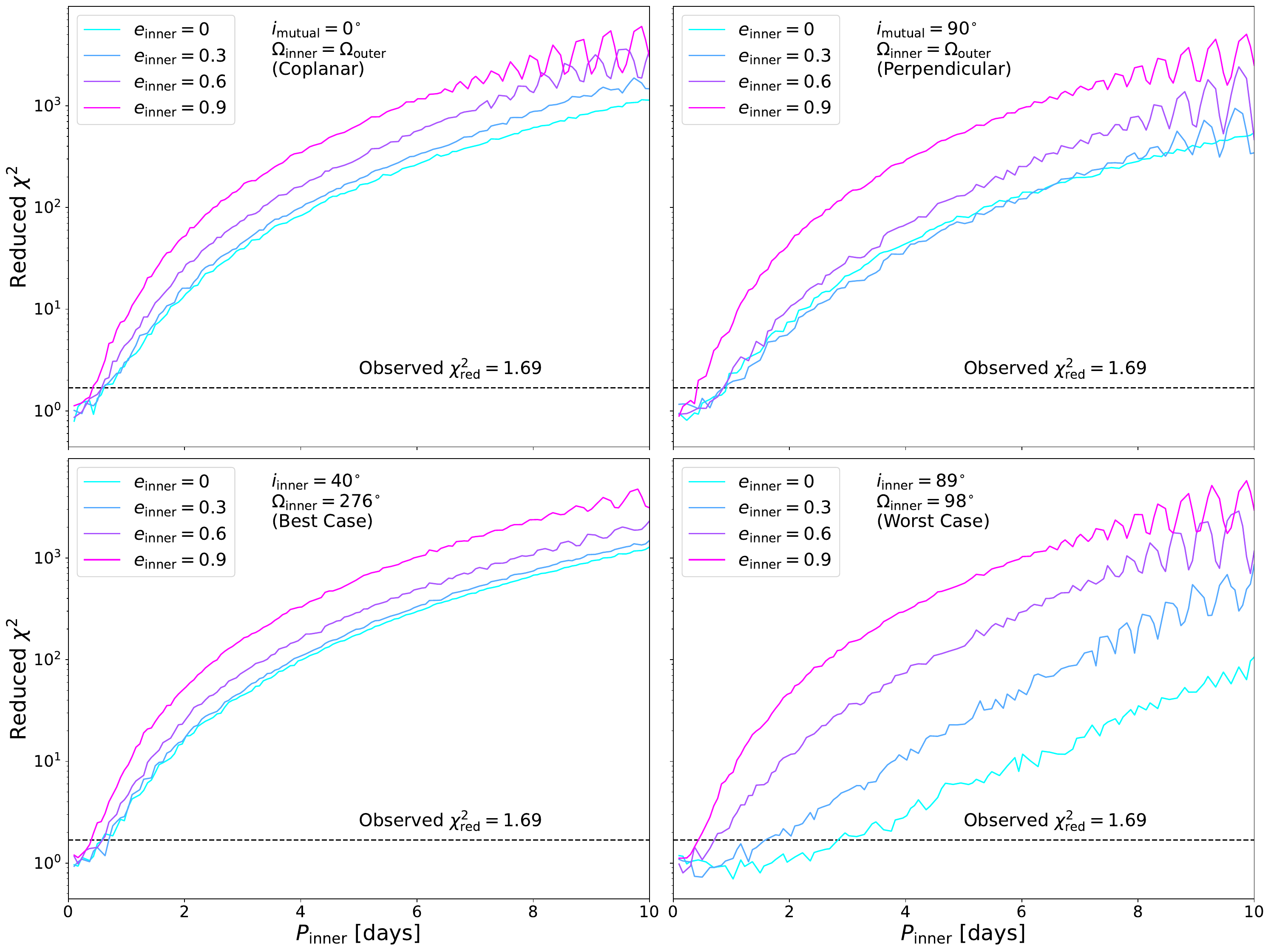}
\caption{Reduced $\chi^2$ value of the best-fit Keplerian model as a function of $P_{\text{inner}}$ for simulated data assuming the exact observing cadence and uncertainties of our measured RVs. Different panels show different orientations of the inner binary, and lines within each panel show different inner binary eccentricities. The reduced $\chi^2$ value generally increases with increasing $e_{\text{inner}}$. At very high $e_{\text{inner}}$, the non-uniform sampling of our RVs gives rise to oscillations in the reduced $\chi^2$ value. Based on our best-fit model's reduced $\chi^2$ value of $1.69$, we can rule out most inner binaries with $P_{\text{inner}} \gtrsim 1.5$ days. Fine-tuned orientations exist that could hide inner binaries with orbital periods up to $\sim 3$ days (e.g. lower right); we explore these in more detail in Figure~\ref{fig:heatmap}.}
\label{fig:detect_bbh}
\end{figure*}

To explore limits on the period of the inner binary (if indeed it exists), we present the reduced $\chi^2$ value of a $\texttt{RadVel}$ Keplerian fit to simulated RVs for a hierarchical triple as a function of $P_{\text{inner}}$ for various orbital configurations of the inner binary in Figure \ref{fig:detect_bbh}. In contrast to our simulations in Section \ref{sec:theory}, these simulations assume the exact observing cadence and uncertainties of our measured RVs. 

As already demonstrated in Figure \ref{fig:all_power_laws}, the predicted RV signal of an inner binary depends on its period, eccentricity, mass ratio, and orientation. In Figure~\ref{fig:detect_bbh}, we provide predictions for four different inner binary orientations, with each panel representing a different combination of inclination and longitude of the ascending node. We show a range of periods and eccentricities for the inner binary in each panel, setting the argument of periastron and periastron time to zero and the mass ratio to unity. We also plot a black dashed line indicating our observed value of reduced $\chi^2 = 1.69$ from the updated Keplerian fit for Gaia BH1. 

For most of the orientations and eccentricities shown in Figure \ref{fig:detect_bbh}, the predicted reduced $\chi^2$ rises above the observed value for $P_{\text{inner}} \gtrsim 1$ day, indicating that we can rule out an inner BBH with $P_{\text{inner}} \gtrsim 1$ day for such orientations. However, for some inner binary orientations -- such as the one labeled ``worst case'' in the bottom right panel -- the predicted RV signature of an inner binary is weaker, only rising above reduced $\chi^2 = 1.69$ for $P_{\text{inner}} \gtrsim 3$ days. 

\begin{figure*}
\epsscale{1.1}
\plotone{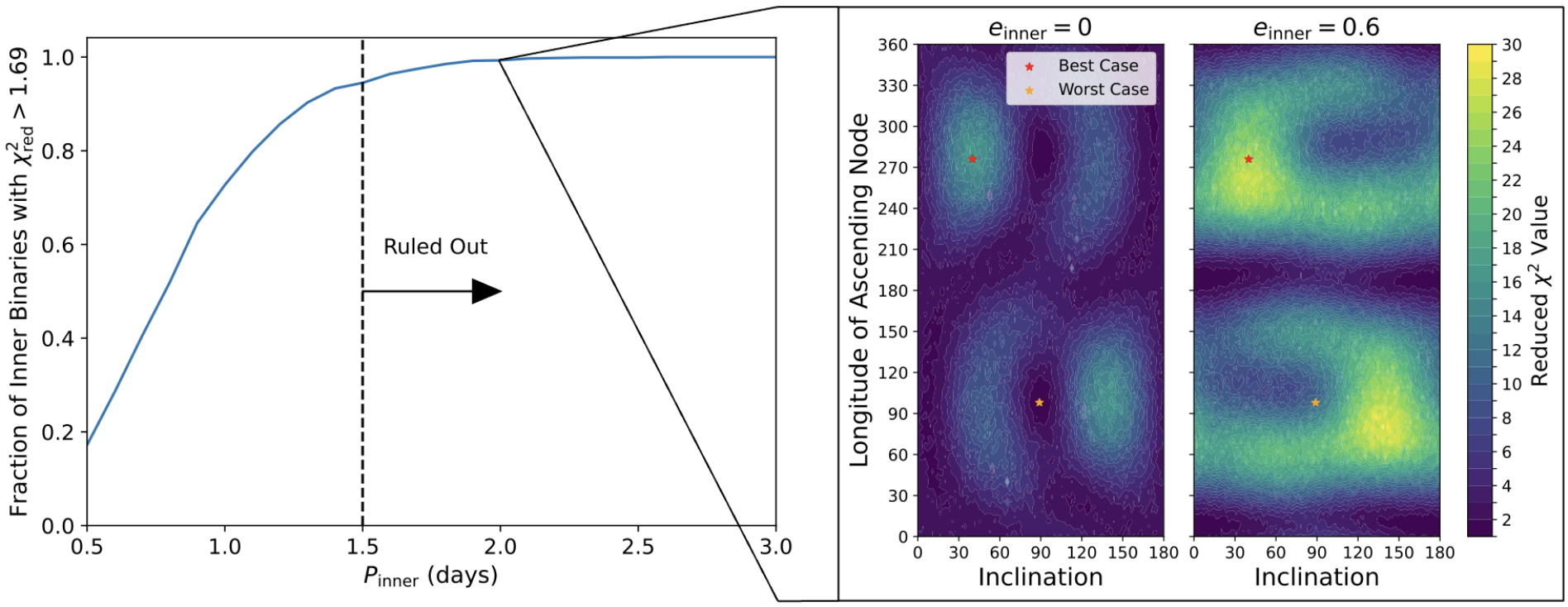}
\caption{Left: Fraction of simulated inner binaries that are detectable with our data as a function of inner period.  We assume random inner binary orientations and a uniform inner binary eccentricity distribution, and we define ``detectable'' binaries as those which produce a worse Keplerian two-body fit than we observe. The observed RVs rule out most ($\gtrsim 95\%$) inner binaries with $P_{\text{inner}} > 1.5$ days. Right: Dependence of the best-fit Keplerian model's reduced $\chi^2$ on the orientation of the inner BBH, for two values of the inner binary eccentricity. For each choice of inner binary inclination and longitude of the ascending node, we simulate observations of a triple with the exact observing cadence and uncertainties of our measured RVs and then fit a Keplerian two-body orbit. We assume $P_{\text{inner}} = 2$ days, as might be expected for an inner BBH just above our detection threshold. The ``best'' and ``worst'' cases shown in Figure \ref{fig:detect_bbh} are labeled. The reduced $\chi^2$ landscape is complex, but only a few fine-tuned inner binary orientations allow reduced $\chi^2 < 2$.}
\label{fig:heatmap}
\end{figure*}

We explore the sensitivity of our constraints to the orientation of the inner binary more thoroughly in the right panels of Figure \ref{fig:heatmap}, where we plot the reduced $\chi^2$ value of the best-fit Keplerian model as a function of inclination and longitude of the ascending node of the inner orbit. We assume $q_{\text{inner}}=1$ and $P_{\text{inner}} = 2$ days in this exploration, as might be expected for an inner binary just above our detection threshold. As before, we adopt the exact observing cadence and uncertainties of our measured RVs, and set the argument of periastron and periastron time of the inner orbit to zero. We show predictions for two values of the inner binary's eccentricity, $e_{\text{inner}}=0$ and $e_{\text{inner}} = 0.6$. We also label the orientations corresponding to the ``best'' and ``worst'' cases (in terms of detectability, for $e_{\text{inner}}=0$) displayed in Figure \ref{fig:detect_bbh}. 

The reduced $\chi^2$ landscape shows the expected rotational and reflection symmetries. Beyond this, it is complicated, bearing imprints of both the observational properties (i.e.\ cadence and uncertainties) of the RVs, and the orientation of the outer orbit relative to our line of sight. For both circular and eccentric inner orbits, the most easily detectable inner binary is close to (but not exactly) coplanar with the outer orbit, while the hardest-to-detect orientations are closer to perpendicular. 

Given the complicated dependence on inner binary eccentricity, mass ratio, and orientation, it is difficult to determine a single upper limit on the inner binary's period. For the most fine-tuned orientations, periods as long as $P_{\text{inner}} \sim 3$ days could escape detection in our data with an equal-mass inner binary, and even longer inner periods are possible if the inner binary mass ratio is allowed to be arbitrarily large. To probe further, we generate $1000$ random eccentricities, inclinations, and longitudes of the ascending node at each value of $P_{\text{inner}}$ on a grid between $0.5$ and $3$ days. We set the argument of periastron and periastron time of the inner orbit to zero, and calculate the reduced $\chi^2$ value in each case, assuming our observed cadence and RV uncertainties. We assume a uniform $e_{\text{inner}}$ distribution and random orientations. Then, at each value of $P_{\text{inner}}$, we compute the fraction of inner binaries that would have been detected at our threshold of reduced $\chi^2 = 1.69$. We plot the resulting curve in the left panel of Figure \ref{fig:heatmap}. We conclude that, for {\it typical} inner binary orientations and eccentric inner orbits -- as are expected in the presence of natal kicks and Kozai-Lidov oscillations (\citealt{1962P&SS....9..719L}, \citealt{1962AJ.....67..591K}, \citealt{naoz_review_2016}) -- our observed RVs rule out most inner binaries with $P_{\text{inner}} \gtrsim 1.5$ days. This is consistent with the best-fit derived value of $P_{\text{inner}} = 0.9 \pm 0.1$ days from our hierarchical triple model. 

\subsection{Comparison of the best-fit binary and triple models}
\begin{figure*}
\epsscale{0.9}
\plotone{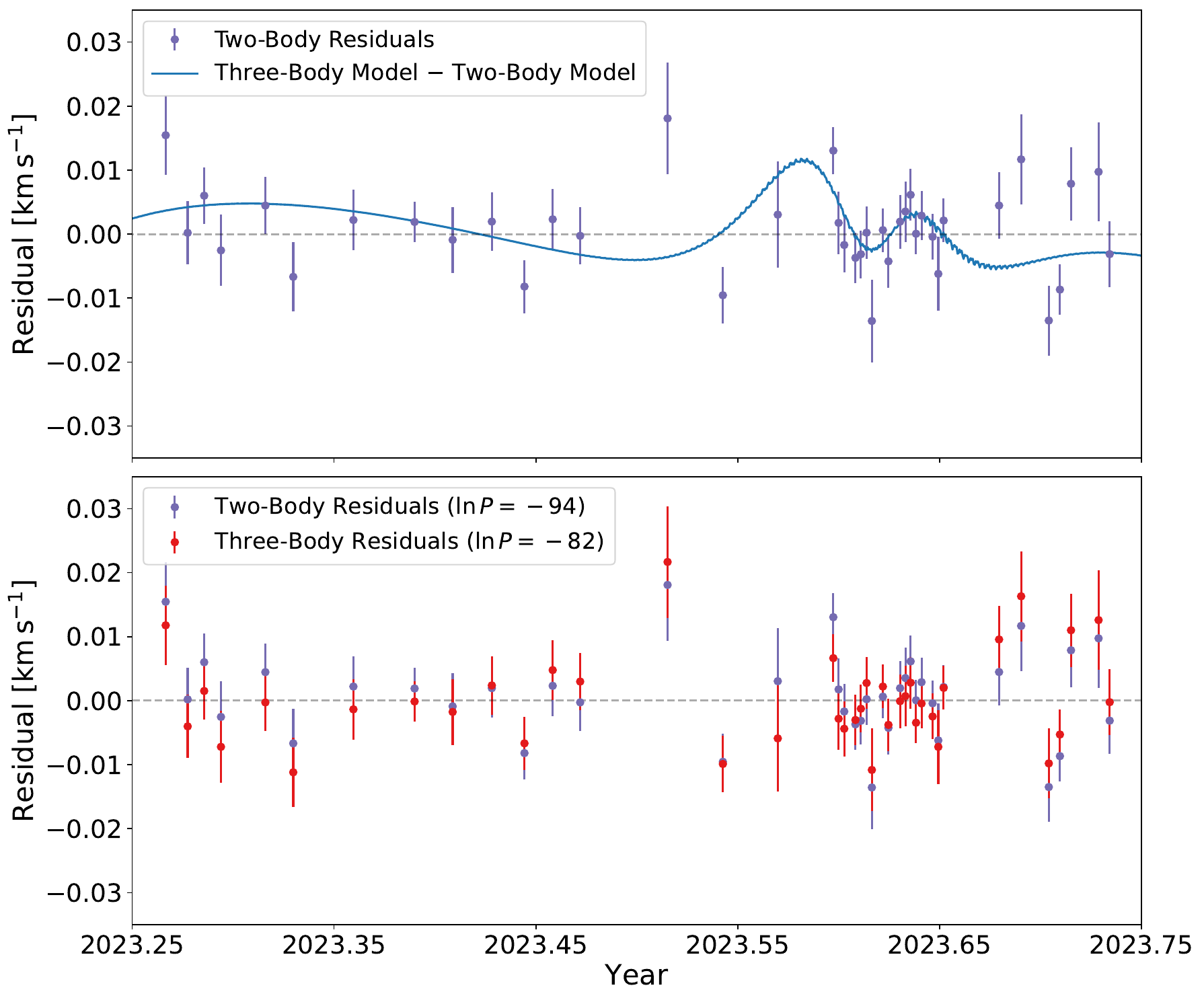}
\caption{Top: Points with error bars show the observed ESPRESSO RV residuals relative to the best-fit two-body Keplerian model. The continuous curve represents the difference between the best-fit hierarchical triple model and that Keplerian model. The residuals are consistent with the continuous curve to within 1--2$\sigma$. Bottom: Observed ESPRESSO RV residuals relative to the best-fit two-body Keplerian model (purple; same as top panel) and the best-fit hierarchical triple model (red). The scatter of the residuals for the hierarchical triple model is somewhat smaller than in the Keplerian case, with an improvement in log posterior probability of $\sim$12. This reflects the fact that the more flexible model allows for a long-term trend in the residuals due to precession. However, the improvement is small compared to the RV uncertainties and appears consistent with overfitting.}
\label{fig:models_residuals}
\end{figure*}

We provide a comparison of the observed RV residuals relative to the best-fit two-body Keplerian model with the difference between the best-fit hierarchical triple and two-body Keplerian models in the top panel of Figure \ref{fig:models_residuals}, focusing on the high-precision ESPRESSO data. We find that the residuals are consistent with the continuous curve representing the difference between our best-fit models to within 1--2$\sigma$. 

We compare observed RV residuals for the best-fit two-body Keplerian and hierarchical triple models in the bottom panel of Figure \ref{fig:models_residuals}, once again displaying only the high-precision ESPRESSO data. The log posterior probability (which includes the RVs from all instruments, the {\it Gaia} astrometry, and the priors)  for the best-fit three body model is $-82$, which is slightly favorable compared to the log posterior probability for the best-fit two-body model of $-94$. While the scatter of the ESPRESSO RV residuals around zero for the hierarchical triple model appears to be slightly smaller than in the Keplerian case, the observed amplitude of the residuals is small, and the improvement is not significant with respect to the RV uncertainties.

From Figure \ref{fig:models_residuals}, we conclude that the hierarchical triple model provides a marginally better fit to the observed RVs compared to the two-body Keplerian model. This is because the extra model parameters allow for a long-term variation (due to precession) that can absorb some of the scatter in the residuals. However, this improvement is small compared to the individual uncertainties; at this level of precision, the difference in the log posterior probability is most likely due to overfitting the noise in the observed data.

To emphasize this, we turn to the Bayesian information criterion (BIC), a model selection metric that penalizes increasing model complexity \citep{schwarz}. The BIC is defined to be:

\begin{equation}
    \text{BIC} = k \ln{n} - 2 \ln{\hat{L}}
\end{equation}

where $k$ is the number of model parameters, $n$ is the number of data points, and $\hat{L}$ is the maximum value of the likelihood function for the model in question. We find that the two-body Keplerian model and hierarchical triple models have BIC values of $246$ and $258$ respectively. Since lower BIC values are generally preferred, the evidence appears to favor the two-body model over the three-body one.

\subsection{BH binary inspiral time and detectability with LISA}
\label{sec:merger_time}
\begin{figure*}
\epsscale{0.85}
\plotone{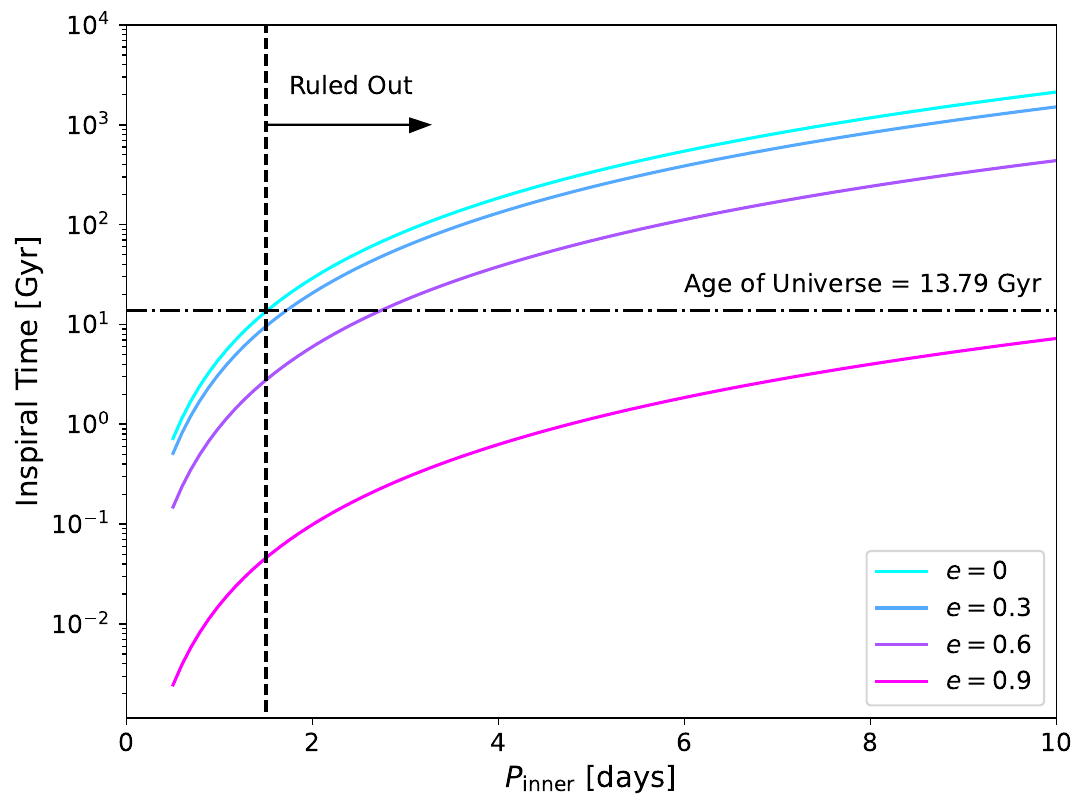}
\caption{Inspiral times of a equal mass ratio black hole binary ($M_{\text{BH}, 1} = M_{\text{BH}, 2} = 4.65 ~ M_{\odot}$) as a function of orbital period for various eccentricities. The BH binary merges faster for higher orbital eccentricities. The fact that equal-mass BH binaries with $P_{\text{inner}} < 1.5$ days merge within a Hubble time implies that hierarchical triple models for Gaia BH1 with inner BH binaries at these orbital periods (though consistent with the observed RV residuals from our two-body Keplerian fit) require some fine tuning.}
\label{fig:inspiral_times}
\end{figure*}

While an inner BH binary with $P_{\text{inner}} < 1.5$ days could possibly exist in $\text{Gaia}$ BH1, this would be a fine-tuned scenario, as such a close binary would have merged in less than a Hubble time. To demonstrate this, we use the results derived by \citet{PhysRev.136.B1224} to calculate the inspiral time for an equal-mass inner BH binary as a function of orbital period at various eccentricities in Figure \ref{fig:inspiral_times}. These calculations account for evolution of the eccentricity (i.e.\ circularization) during the inspiral. We make no attempt to account for perturbations due to the outer star, which would accelerate the merger via Kozai-Lidov oscillations for some orbital configurations. Based on our best-fit models, we choose a total binary BH mass of $9.3$ solar masses. We confirm that the BBH inspiral time is shorter at higher orbital eccentricities. Furthermore, at $P_{\text{inner}} < 1.5$ days, the BH binary merges in less than the age of the Universe, even for cases with low orbital eccentricities.

At sufficiently short periods or high eccentricities, an inner BH binary would be detectable via gravitational waves (GWs) with the \textit{LISA} space observatory \citep{LISA_2017}. We use \texttt{LEGWORK} \citep{LEGWORK_apjs} to estimate the SNR of an equal-mass BH binary in Gaia BH1 observed by {\it LISA} for a range of inner periods and eccentricities, assuming a 4-year mission duration. We find that binaries with $P_{\text{inner}} \lesssim 0.15$ days would be detectable with SNR $>$ 5 for any eccentricity. For the same SNR threshold, the maximum detectable inner orbital period rises to $0.33$ days for $e_{\text{inner}} = 0.5$ and $2.5$ days for $e_{\text{inner}} = 0.9$. At higher eccentricities, the detection is primarily due to the GW signal from higher harmonics of the orbital period. Figure \ref{fig:inspiral_times} demonstrates that catching any individual BH binary at a period where it would be detectable by {\it LISA} but had not already merged requires some fine tuning. Of course, {\it LISA} will have the advantage of being sensitive to {\it all} close BH binaries in the Milky Way, including those not orbited by a luminous tertiary.

\subsection{Limits on the presence of distant tertiary star}
\label{sec:dist_tertiary}

\begin{figure*}
\epsscale{0.85}
\plotone{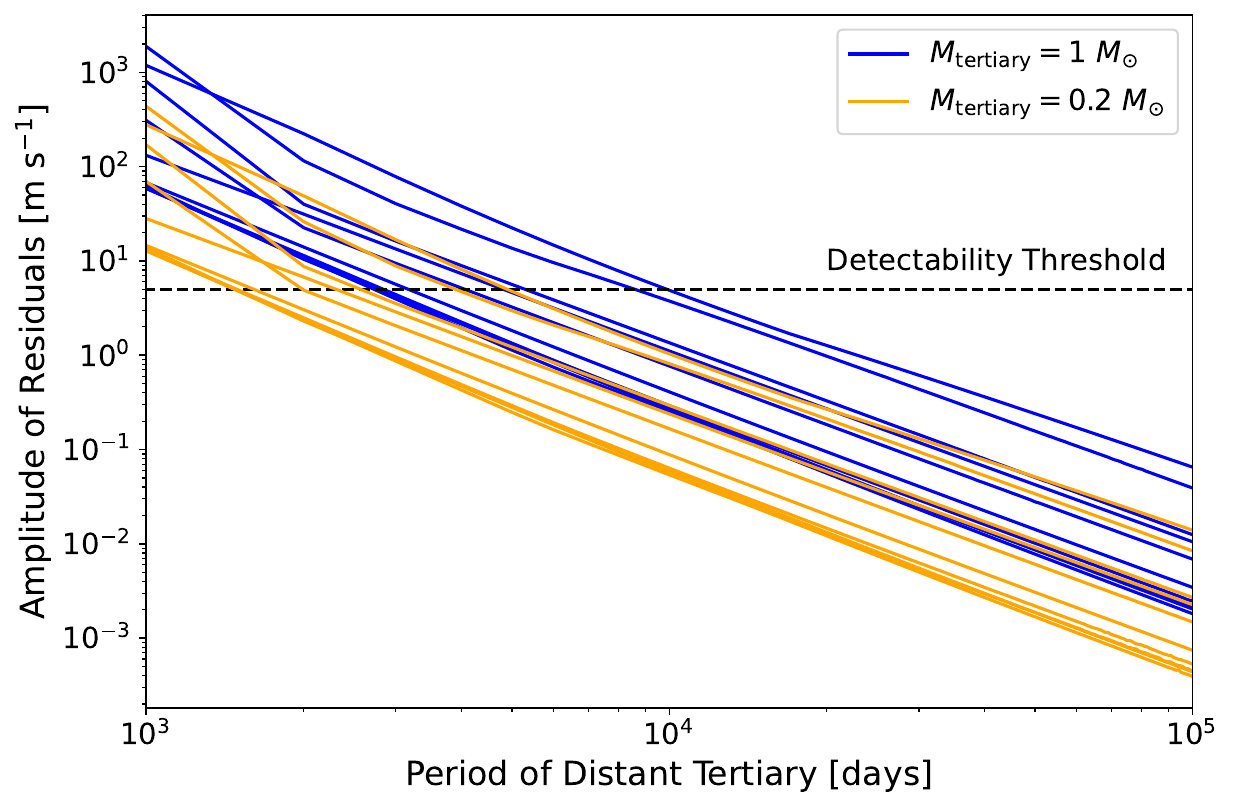}
\caption{RV residual amplitudes predicted for the Sun-like star when the star + BH binary is orbited by a distant (outer) tertiary. Blue and orange curves correspond to a $1 ~ M_{\odot}$ white dwarf and a $0.2 ~ M_{\odot}$ M dwarf, respectively, with high-precision RVs obtained over one orbit. Different curves show different random orbital phases. Based on our detectability threshold of an RV residual amplitude of $\sim$ 5 m s$^{-1}$, we can rule out a tertiary white dwarf at all orbital periods less than about $2700$ days, and a tertiary M dwarf at periods less than about $1500$ days.}
\label{fig:detect_teritary}
\end{figure*}

The fact that our RVs are well-fit by a Keplerian two-body orbit also places constraints on the presence of objects in wider orbits in the Gaia BH1 system. We investigate the perturbation that would result from a distant tertiary star by plotting the Keplerian RV residual amplitude of the Sun-like star as a function of orbital period of the tertiary in Figure \ref{fig:detect_teritary}. We consider the cases of $1 ~ M_{\odot}$ white dwarf tertiary and a $0.2 ~ M_{\odot}$ M dwarf tertiary, both of which could have evaded spectroscopic detection. For the tertiary star, we adopt an orbital eccentricity of $0.5$ and set both the inclination and longitude of the ascending node to zero. In addition, we assume the same observing time baseline as in Figure \ref{fig:intro_fig}. For each tertiary, we randomly sample ten orbital phases and plot the resulting curves in Figure \ref{fig:detect_teritary}. 

Adopting a detectability threshold of a RV residual amplitude of $\sim 5$ m s$^{-1}$, we find that we would detect a $1 ~ M_{\odot}$ tertiary at orbital periods less than about $2700$ days, or approximately $7$ years, in the residuals of a two-body Keplerian fit. For a $0.2 ~ M_{\odot}$ tertiary, the lower limit on the orbital period is about $1500$ days, or approximately $4$ years, instead. We conclude that sensitivity to a low-mass tertiary star would require spectroscopic follow-up over an extremely long observational baseline.

Tertiaries that are bright enough to be detected by {\it Gaia} (corresponding to masses $\gtrsim 0.2 ~ M_{\odot}$ for main-sequence stars at the distance of Gaia BH1) are already ruled out. Assuming a 1 arcsec {\it Gaia} resolution (which is somewhat optimistic for the faintest companions due to blending; \citealt{El-Badry2018}), this rules out separations wider than $\sim 500$ AU, or periods $<10^6$ days. A stronger limit of $\approx 0.1$ arcsec, corresponding to periods $\lesssim 10^{4.5}$ days, could be set with speckle interferometry \citep[e.g.][]{2022FrASS...9.1163H}.

\subsection{Implications for the formation history of Gaia BH1}
\label{sec:formation_history}

We have shown that the $\sim 9.3 ~ M_{\odot}$ dark object in the Gaia BH1 system is unlikely to be a binary with period longer than $\sim 1.5$ days. Shorter inner periods cannot be ruled out because they would lead to deviations from a Keplerian orbit that could not be detected with the current data. However, such short inner binaries would merge within a Hubble time and thus require significant fine tuning. Hence, the simplest --- and arguably most plausible --- explanation is that the dark object is a single, $9.3 ~ M_{\odot}$ BH. 

It is in principle possible that the dark object was initially a BH + BH or BH + NS binary and has already merged. However, we consider this scenario unlikely because anisotropic gravitational wave emission produces a kick during compact object mergers, with a typical magnitude above $100$ km s$^{-1}$ \citep[e.g.][]{Bekenstein1973, Merritt2004}. Such a kick would most likely have disrupted the outer binary, or imparted a larger eccentricity and space velocity on it than is observed. It is also possible that the system started as a triple but the inner binary merged before one or both stars died, as some evolutionary models predict merger products to explode as blue supergiants or Wolf-Rayet stars without ever expanding to red supergiant dimensions \citep[e.g.][]{Justham2014}. The range of initial triple conditions for which such an evolutionary scenario is feasible in the Gaia BH1 system is, however, likely rather limited. A related  possibility is that the BH could have formed from a progenitor with mass $\gtrsim 40 ~ M_{\odot}$ that never became a red supergiant \citep[e.g.][]{HD_1979, davies_2018, gilkis_2021}, but it is still uncertain whether very massive stars briefly expand to red supergiant dimensions in their post-main sequence evolution.

Another formation channel that has recently been explored is dynamical assembly in a star cluster. The G star's near-solar metallicity and disk-like orbit disfavor a globular cluster, but the system may have formed in a cluster in the Galactic disk that has since dissolved. This scenario is difficult to test because it makes no predictions for present-day observables that could distinguish a dynamically-assembled BH binary from a primordial one. Several recent works have predicted that dynamical formation in intermediate- or high-mass clusters could dominate the formation rate of Gaia BH1-like binaries \citep{rastello2023dynamical, dicarlo2023young, Tanikawa2023}. However, the models considered so far (a) appear rather fine-tuned, involving multiple protagonists, stellar mergers, and a calibrated BH natal kick, (b) do not actually avoid a common envelope event, which the binary may not survive, and (c) predict binaries with periods $P_{\text{orb}} \sim 100$ days to form inefficiently compared to both shorter and longer periods \citep{rastello2023dynamical}. Further work is required to explore the sensitivity of these predictions to initial cluster mass and primordial binary population. 

Another possibility is that Gaia BH1 formed through isolated binary evolution via a channel not captured in vanilla population synthesis models \citep[e.g.][]{Hirai2022}. The large populations of wide white dwarf + main sequence  \citep{Yamaguchi2023, Shahaf2023} and neutron star + main sequence (El-Badry et al., in prep) binaries discovered by {\it Gaia} with orbits similar to that of Gaia BH1 lends some credibility to this possibility, but more work is required to explore it.

\subsection{Can further spectroscopic follow-up detect a inner BH binary in Gaia BH1?}

As shown in Section \ref{sec:baseline}, sensitivity to an inner binary improves significantly as the observing time baseline increases. This is primarily because inner binaries cause the outer orbit to precess, and the cumulative effects of precession grow over time. 

Extending our analysis in Section \ref{sec:limits}, we find that acquiring 20 additional ESPRESSO RVs of Gaia BH1 near the August periastron passages in 2024 and 2025 would improve our constraint on the orbital period of an inner BBH to an upper limit of about $0.75$ days. In this work, our best-fit hierarchical triple model for Gaia BH1 suggests the possibility of an inner BH binary with $P_{\text{inner}} = 0.9 \pm 0.1$ days. Thus, future work on high-precision spectroscopic follow-up of Gaia BH1 would already be able to rule out (or confirm) our highest-probability hierarchical triple solution. While improving the upper limit of $P_{\text{inner}} \lesssim 1.5$ days to $P_{\text{inner}} \lesssim 0.75$ days may seem incremental, this would reduce the implied minimum inspiral time from the age of the Universe to only $2$ Gyr (see Figure \ref{fig:inspiral_times}).

\section{Conclusions} \label{sec:conclusion}

We have presented high-precision RV follow-up of Gaia BH1, a nearby binary system containing a Sun-like star orbiting a dark object with mass $\sim 9.3\,M_{\odot}$. The system was recently discovered via {\it Gaia} astrometry and further characterized with low-precision RV follow-up. In their discovery paper, \citetalias{Gaia_BH1} interpreted the dark object as a likely dormant stellar-mass black hole (BH). They noted, however, that their data were also consistent with it being a close BH + BH or BH + NS binary, and that a hierarchical triple origin for the system could potentially alleviate some of the challenges associated with explaining its formation.  

If Gaia BH1 hosts an inner black hole binary (BBH), the luminous star's RVs will display subtle deviations from a Keplerian orbit. High-precision RVs can thus put the hierarchical triple model to the test. In parallel with our work, the feasibility of such a test was recently explored by \citet{hayashi_suto_trani_2023} using idealized simulations. Here, we report new RV data and simulations matched to that data. Our main conclusions are as follows: 

\begin{itemize}
    \item {\it Sensitivity to inner BH binaries:} We explore the parameter space of expected RV signatures of a hierarchical triple in the non-circular, non-coplanar regime using the N-body integrator $\texttt{REBOUND}$ (Figure \ref{fig:intro_fig}). We find that (a) increasing the eccentricity of either the inner or outer orbit increases the amplitude of observed RV residuals, (b) increasing the inner binary mass ratio decreases the amplitude of observed RV residuals, and (c) varying the  mutual inclination non-monotonically changes the amplitude of observed RV residuals  (see Figure \ref{fig:all_power_laws}). We determine that with optimal observing cadence we can reasonably detect an inner BH binary in the RV residuals of a Keplerian fit for inner periods $\gtrsim$ 1 day (see Figure \ref{fig:realistic_fig}), and that residuals in realistic orbital configurations are always larger than in the circular and coplanar case (Figure \ref{fig:typical_fig}).
    
    We show that with observations over one period of the outer orbit, the short-timescale RV variations induced by an inner BH binary are of comparable magnitude to the long-timescale RV variations due to precession (see Figure \ref{fig:period_fig}). Extending the observational baseline to multiple orbital cycles helps detect the cumulative effect of precession in the outer star's orbit (Figure \ref{fig:cycle_fig}).
    
    We find that it is possible to distinguish between perturbations due to an inner BH binary and an exoplanet orbiting the outer star with precise RV measurements. A clear difference between the two cases is that the amplitude of RV variations does not increase significantly at the periastron of the outer star's orbit in the exoplanetary case (last panel of Figure \ref{fig:typical_fig}).
    
    \item {\it High-precision RV data:} We obtained 40 high-precision RV measurements ($\sim 3$--$5$ m s$^{-1}$ uncertainties) using ESPRESSO over one orbit of the Sun-like star. We supplement these data with 75 (mostly new) medium-precision RVs  ($\sim 30$--$100$ m s$^{-1}$ uncertainties) measured with FEROS, TRES, and HIRES over three orbits (Figure \ref{fig:updated_astro_fig} and Table~\ref{tab:all_rvs}). We concentrated the high-precision RVs near a periastron passage, where the expected perturbations due to an inner binary are largest. We combine the spectroscopic and astrometric data to tightly constrain the parameters of a two-body Keplerian model and update the constraints from \citetalias{Gaia_BH1} (Table \ref{tab:gaia_bh1_params} and Figure \ref{fig:overlaid_corner_plot}). 
    
    \item {\it Two-body and three-body fits:} The observed RVs are well-fit by a Keplerian two-body orbit. From our observed value of reduced $\chi^2 = 1.69$, we set an upper limit of $P_{\text{inner}} \lesssim 1.5$ days on the period of any inner BBH (see Figures \ref{fig:detect_bbh} and \ref{fig:heatmap}). 
    
    To assess whether a model with an inner BH binary can better explain the data than a two-body Keplerian model, we use \texttt{REBOUND} to fit a three-body model. We derive the best-fit orbital parameters for both the outer star and the inner BH binary in this case (see Table \ref{tab:gaia_bh1_triple}). We find that the putative inner BH binary would have an orbital period of $0.9 \pm 0.1$ days, consistent with our limits from the Keplerian fit. While the hierarchical triple model can provide a marginally better fit to the observed RVs than the two-body Keplerian model (Figure \ref{fig:models_residuals}), the improvement is not sufficient to justify the model's higher complexity, and we conclude that  the improvement in log posterior probability is most likely due to overfitting the noise.

    Acquiring $\sim20$ additional precise RV observations of Gaia BH1 around periastron in 2024 and 2025 would improve our constraint on the orbital period of the inner BBH to an upper limit of about $0.75$ days. 
    
    \item {\it Merger timescale:} While it is possible that an inner BH binary with $P_{\text{inner}} < 1.5$ days exists, such a BH binary would be expected to merge in less than a Hubble time (see Figure \ref{fig:inspiral_times}). Hiding a BH binary in the period range not ruled out by RVs would thus require fine-tuning. Much of the short-period inner binary regime not ruled out by our RVs will eventually be probed by {\it LISA} (Section \ref{sec:merger_time}).
    
    \item {\it Implications for the formation of Gaia BH1:} Our results imply that Gaia BH1 is unlikely to currently host an inner BH binary: a single $9.3\,M_{\odot}$ BH is the most likely companion. It is also unlikely that it previously hosted such a binary that has now merged, because the kick due to anisotropic gravitational wave emission would likely have unbound the outer orbit. We have {\it not} ruled out a triple scenario in which the inner binary merged while its components were still on the main sequence. More work is required to investigate alternative formation channels, such as dynamical assembly through exchange interactions, nonstandard treatments of common envelope evolution, and efficient wind mass loss to avoid a red supergiant phase in the BH progenitor. 
\end{itemize}


\begin{acknowledgments}

We thank Toshinori Hayashi, Jim Fuller, and Dave Charbonneau for helpful discussions. PN and KE were supported in part by NSF grant AST-2307232. AT and TB are supported by a grant from the European Research Council (ERC) under the European Union’s Horizon 2020 research and innovation program (grant agreement number 803193/BEBOP). HWR acknowledges the European Research Council for support from the ERC Advanced Grant ERC-2021-ADG-101054731.
\end{acknowledgments}

%

\vspace{5mm}
\facilities{ESPRESSO (VLT)}


\software{astropy \citep{2013A&A...558A..33A,2018AJ....156..123A}, \texttt{REBOUND} \citep{rebound_zenodo}, \texttt{RadVel} \citep{radvel_zenodo}, \texttt{emcee} \citep{emcee_2013}, \texttt{dynesty} \citep{koposov_2023}}, \texttt{LEGWORK} \citep{LEGWORK_joss}



\appendix

\section{Radial Velocity Data}

We present all radial velocity measurements for Gaia BH1 used in this work in chronological order in Table \ref{tab:all_rvs} below. The MagE, GMOS, XSHOOTER, and ESI data are taken from \citetalias{Gaia_BH1}. The remainder of the data (from the HIRES, FEROS, TRES, and ESPRESSO instruments) are either updated in or new to this work.

\startlongtable
\begin{deluxetable*}{cccc}
\tablecaption{All RV measurements obtained via spectroscopic follow-up of Gaia BH1 using the MagE, GMOS, XSHOOTER, ESI, HIRES, FEROS, TRES, and ESPRESSO instruments. \label{tab:all_rvs}}
\tablehead{\colhead{HJD} & \colhead{Radial Velocity (km s$^{-1}$)} & \colhead{Instrument} \\
\colhead{(1)} & \colhead{(2)} & \colhead{(3)}}
\startdata
2459767.6226 &          $63.8 \pm 3$ &       MagE \\
2459791.9186 &      $131.90 \pm 0.1$ &      HIRES \\
2459795.6461 &         $141.4 \pm 3$ &       MagE \\
2459796.4995 &         $142.7 \pm 3$ &       MagE \\
2459798.8399 &         $140.6 \pm 4$ &       GMOS \\
2459805.5101 &       $127.7 \pm 1.0$ &   XSHOOTER \\
2459808.7388 &         $118.0 \pm 4$ &       GMOS \\
2459813.6045 &      $90.47 \pm 0.15$ &      FEROS \\
2459814.5874 &      $86.23 \pm 0.10$ &      FEROS \\
2459815.5927 &    $82.090 \pm 0.049$ &      FEROS \\
2459817.5278 &    $74.475 \pm 0.053$ &      FEROS \\
2459818.5266 &    $70.810 \pm 0.054$ &      FEROS \\
2459818.7870 &          $67.8 \pm 4$ &       GMOS \\
2459819.5543 &    $67.205 \pm 0.058$ &      FEROS \\
2459820.5465 &    $63.950 \pm 0.067$ &      FEROS \\
2459821.5669 &    $60.685 \pm 0.059$ &      FEROS \\
2459822.5745 &    $57.665 \pm 0.085$ &      FEROS \\
2459823.5422 &      $54.98 \pm 0.11$ &      FEROS \\
2459823.8525 &       $53.76 \pm 0.1$ &      HIRES \\
2459824.5306 &    $52.200 \pm 0.033$ &      FEROS \\
2459824.8516 &       $51.18 \pm 0.1$ &      HIRES \\
2459825.5361 &      $49.90 \pm 0.16$ &      FEROS \\
2459826.7920 &       $46.59 \pm 0.1$ &      HIRES \\
2459828.5677 &    $42.795 \pm 0.073$ &      FEROS \\
2459829.5373 &          $42.1 \pm 3$ &       MagE \\
2459829.5768 &    $40.520 \pm 0.086$ &      FEROS \\
2459830.6175 &      $38.76 \pm 0.15$ &      FEROS \\
2459831.6223 &      $37.34 \pm 0.38$ &      FEROS \\
2459833.7523 &       $33.23 \pm 0.1$ &      HIRES \\
2459834.5509 &      $31.81 \pm 0.14$ &      FEROS \\
2459834.7691 &       $31.74 \pm 0.1$ &      HIRES \\
2459835.7678 &       $30.14 \pm 0.1$ &      HIRES \\
2459838.7208 &          $27.5 \pm 4$ &       GMOS \\
2459838.8082 &       $26.35 \pm 0.1$ &      HIRES \\
2459840.7729 &       $24.20 \pm 0.1$ &      HIRES \\
2459845.5069 &        $19.4 \pm 1.0$ &   XSHOOTER \\
2459855.5012 &        $14.2 \pm 1.0$ &   XSHOOTER \\
2459868.5128 &         $9.3 \pm 1.0$ &   XSHOOTER \\
2459877.6978 &        $10.5 \pm 1.5$ &        ESI \\
2459985.8801 &     $139.05 \pm 0.23$ &      FEROS \\
2459989.8698 &     $129.35 \pm 0.08$ &      FEROS \\
2459992.8690 &     $117.79 \pm 0.06$ &      FEROS \\
2459996.8740 &      $99.86 \pm 0.09$ &      FEROS \\
2459997.8709 &      $95.49 \pm 0.15$ &      FEROS \\
2459999.8760 &      $86.66 \pm 0.09$ &      FEROS \\
2460000.8793 &      $82.43 \pm 0.07$ &      FEROS \\
2460001.8731 &      $78.57 \pm 0.11$ &      FEROS \\
2460002.8546 &      $74.67 \pm 0.08$ &      FEROS \\
2460003.8613 &      $71.13 \pm 0.06$ &      FEROS \\
2460004.8639 &      $67.51 \pm 0.07$ &      FEROS \\
2460005.8594 &      $64.21 \pm 0.04$ &      FEROS \\
2460006.8705 &      $60.87 \pm 0.05$ &      FEROS \\
2460012.8827 &      $44.99 \pm 0.04$ &      FEROS \\
2460014.8826 &      $40.84 \pm 0.07$ &      FEROS \\
2460016.8810 &      $37.03 \pm 0.05$ &      FEROS \\
2460017.8634 &      $35.43 \pm 0.05$ &      FEROS \\
2460018.8627 &      $33.73 \pm 0.05$ &      FEROS \\
2460026.8575 &      $23.50 \pm 0.07$ &      FEROS \\
2460037.9770 &    $15.670 \pm 0.059$ &       TRES \\
2460040.8413 &      $14.03 \pm 0.05$ &      FEROS \\
2460042.8033 &  $13.0181 \pm 0.0062$ &   ESPRESSO \\
2460043.8640 &      $12.87 \pm 0.05$ &      FEROS \\
2460046.7778 &  $11.7593 \pm 0.0049$ &   ESPRESSO \\
2460049.7651 &  $11.0815 \pm 0.0044$ &   ESPRESSO \\
2460050.8708 &      $11.24 \pm 0.08$ &      FEROS \\
2460050.8927 &    $11.275 \pm 0.056$ &       TRES \\
2460052.8000 &  $10.5756 \pm 0.0056$ &   ESPRESSO \\
2460053.7976 &      $10.60 \pm 0.09$ &      FEROS \\
2460054.8315 &    $10.620 \pm 0.089$ &      FEROS \\
2460060.8260 &  $10.1052 \pm 0.0044$ &   ESPRESSO \\
2460065.8700 &  $10.3429 \pm 0.0054$ &   ESPRESSO \\
2460073.7926 &      $11.75 \pm 0.09$ &      FEROS \\
2460075.8674 &    $12.613 \pm 0.047$ &       TRES \\
2460076.7473 &  $12.1620 \pm 0.0047$ &   ESPRESSO \\
2460076.7645 &    $12.330 \pm 0.066$ &      FEROS \\
2460077.7304 &    $12.725 \pm 0.072$ &      FEROS \\
2460080.8029 &    $13.645 \pm 0.083$ &      FEROS \\
2460085.7878 &    $15.200 \pm 0.068$ &      FEROS \\
2460087.8111 &  $15.7071 \pm 0.0032$ &   ESPRESSO \\
2460089.9047 &    $17.140 \pm 0.049$ &       TRES \\
2460093.8351 &    $18.835 \pm 0.052$ &       TRES \\
2460094.6938 &  $18.8160 \pm 0.0051$ &   ESPRESSO \\
2460101.7831 &  $22.8400 \pm 0.0046$ &   ESPRESSO \\
2460101.8146 &    $23.060 \pm 0.037$ &      FEROS \\
2460102.8134 &    $24.065 \pm 0.046$ &       TRES \\
2460107.6790 &  $26.8994 \pm 0.0042$ &   ESPRESSO \\
2460110.7907 &      $29.47 \pm 0.10$ &      FEROS \\
2460111.7161 &    $30.370 \pm 0.054$ &      FEROS \\
2460112.7846 &  $31.0566 \pm 0.0047$ &   ESPRESSO \\
2460113.6682 &    $32.065 \pm 0.037$ &      FEROS \\
2460113.8126 &    $32.575 \pm 0.059$ &       TRES \\
2460117.7582 &  $35.7466 \pm 0.0045$ &   ESPRESSO \\
2460121.7794 &    $40.755 \pm 0.072$ &       TRES \\
2460133.5611 &  $56.5368 \pm 0.0087$ &   ESPRESSO \\
2460133.7396 &    $57.465 \pm 0.065$ &       TRES \\
2460142.6035 &    $74.505 \pm 0.054$ &      FEROS \\
2460143.5587 &   $76.516 \pm 0.0044$ &   ESPRESSO \\
2460153.4901 & $103.8949 \pm 0.0083$ &   ESPRESSO \\
2460162.6793 &   $132.722 \pm 0.095$ &       TRES \\
2460163.5326 & $134.0932 \pm 0.0037$ &   ESPRESSO \\
2460164.4936 & $136.1711 \pm 0.0049$ &   ESPRESSO \\
2460165.5038 & $138.0140 \pm 0.0043$ &   ESPRESSO \\
2460167.5333 & $140.4032 \pm 0.0039$ &   ESPRESSO \\
2460168.5190 & $140.8277 \pm 0.0037$ &   ESPRESSO \\
2460169.5509 & $140.7025 \pm 0.0041$ &   ESPRESSO \\
2460170.5049 & $140.0269 \pm 0.0065$ &   ESPRESSO \\
2460172.4964 & $136.9670 \pm 0.0034$ &   ESPRESSO \\
2460173.5017 & $134.5784 \pm 0.0041$ &   ESPRESSO \\
2460175.6095 & $128.0143 \pm 0.0042$ &   ESPRESSO \\
2460176.5698 & $124.4401 \pm 0.0047$ &   ESPRESSO \\
2460177.5139 & $120.6559 \pm 0.0040$ &   ESPRESSO \\
2460178.4970 & $116.4871 \pm 0.0031$ &   ESPRESSO \\
2460179.4876 & $112.1362 \pm 0.0038$ &   ESPRESSO \\
2460181.4862 & $103.1152 \pm 0.0036$ &   ESPRESSO \\
2460182.5170 &  $98.4518 \pm 0.0058$ &   ESPRESSO \\
2460183.4844 &  $94.1418 \pm 0.0034$ &   ESPRESSO \\
2460187.6126 &    $77.225 \pm 0.063$ &      FEROS \\
2460192.6551 &    $59.970 \pm 0.042$ &       TRES \\
2460193.5093 &  $56.9746 \pm 0.0052$ &   ESPRESSO \\
2460197.4998 &  $46.5795 \pm 0.0070$ &   ESPRESSO \\
2460202.5055 &  $36.3992 \pm 0.0055$ &   ESPRESSO \\
2460202.6309 &    $36.623 \pm 0.030$ &       TRES \\
2460204.5154 &  $33.0720 \pm 0.0040$ &   ESPRESSO \\
2460206.5202 &  $30.1237 \pm 0.0057$ &   ESPRESSO \\
2460206.6351 &    $30.410 \pm 0.031$ &       TRES \\
2460211.5028 &  $24.0717 \pm 0.0078$ &   ESPRESSO \\
2460213.5108 &  $22.0683 \pm 0.0052$ &   ESPRESSO \\
2460223.5072 &    $15.295 \pm 0.099$ &      FEROS \\
\enddata
\end{deluxetable*}


\bibliography{bibliography}{}

\begin{thebibliography}{}
\expandafter\ifx\csname natexlab\endcsname\relax\def\natexlab#1{#1}\fi
\providecommand{\url}[1]{\href{#1}{#1}}
\providecommand{\dodoi}[1]{doi:~\href{http://doi.org/#1}{\nolinkurl{#1}}}
\providecommand{\doeprint}[1]{\href{http://ascl.net/#1}{\nolinkurl{http://ascl.net/#1}}}
\providecommand{\doarXiv}[1]{\href{https://arxiv.org/abs/#1}{\nolinkurl{https://arxiv.org/abs/#1}}}

\bibitem[{{Aarseth} \& {Mardling}(2001)}]{aarseth_mardling_2001}
{Aarseth}, S.~J., \& {Mardling}, R.~A. 2001, in Astronomical Society of the
  Pacific Conference Series, Vol. 229, Evolution of Binary and Multiple Star
  Systems, ed. P.~{Podsiadlowski}, S.~{Rappaport}, A.~R. {King}, F.~{D'Antona},
  \& L.~{Burderi}, 77, \dodoi{10.48550/arXiv.astro-ph/0011514}

\bibitem[{{Amaro-Seoane} {et~al.}(2017){Amaro-Seoane}, {Audley}, {Babak},
  {Baker}, {Barausse}, {Bender}, {Berti}, {Binetruy}, {Born}, {Bortoluzzi},
  {Camp}, {Caprini}, {Cardoso}, {Colpi}, {Conklin}, {Cornish}, {Cutler},
  {Danzmann}, {Dolesi}, {Ferraioli}, {Ferroni}, {Fitzsimons}, {Gair}, {Gesa
  Bote}, {Giardini}, {Gibert}, {Grimani}, {Halloin}, {Heinzel}, {Hertog},
  {Hewitson}, {Holley-Bockelmann}, {Hollington}, {Hueller}, {Inchauspe},
  {Jetzer}, {Karnesis}, {Killow}, {Klein}, {Klipstein}, {Korsakova}, {Larson},
  {Livas}, {Lloro}, {Man}, {Mance}, {Martino}, {Mateos}, {McKenzie},
  {McWilliams}, {Miller}, {Mueller}, {Nardini}, {Nelemans}, {Nofrarias},
  {Petiteau}, {Pivato}, {Plagnol}, {Porter}, {Reiche}, {Robertson},
  {Robertson}, {Rossi}, {Russano}, {Schutz}, {Sesana}, {Shoemaker}, {Slutsky},
  {Sopuerta}, {Sumner}, {Tamanini}, {Thorpe}, {Troebs}, {Vallisneri},
  {Vecchio}, {Vetrugno}, {Vitale}, {Volonteri}, {Wanner}, {Ward}, {Wass},
  {Weber}, {Ziemer}, \& {Zweifel}}]{LISA_2017}
{Amaro-Seoane}, P., {Audley}, H., {Babak}, S., {et~al.} 2017, arXiv e-prints,
  arXiv:1702.00786, \dodoi{10.48550/arXiv.1702.00786}

\bibitem[{{Astropy Collaboration} {et~al.}(2013){Astropy Collaboration},
  {Robitaille}, {Tollerud}, {Greenfield}, {Droettboom}, {Bray}, {Aldcroft},
  {Davis}, {Ginsburg}, {Price-Whelan}, {Kerzendorf}, {Conley}, {Crighton},
  {Barbary}, {Muna}, {Ferguson}, {Grollier}, {Parikh}, {Nair}, {Unther},
  {Deil}, {Woillez}, {Conseil}, {Kramer}, {Turner}, {Singer}, {Fox}, {Weaver},
  {Zabalza}, {Edwards}, {Azalee Bostroem}, {Burke}, {Casey}, {Crawford},
  {Dencheva}, {Ely}, {Jenness}, {Labrie}, {Lim}, {Pierfederici}, {Pontzen},
  {Ptak}, {Refsdal}, {Servillat}, \& {Streicher}}]{2013A&A...558A..33A}
{Astropy Collaboration}, {Robitaille}, T.~P., {Tollerud}, E.~J., {et~al.} 2013,
  \aap, 558, A33, \dodoi{10.1051/0004-6361/201322068}

\bibitem[{{Astropy Collaboration} {et~al.}(2018){Astropy Collaboration},
  {Price-Whelan}, {Sip{\H{o}}cz}, {G{\"u}nther}, {Lim}, {Crawford}, {Conseil},
  {Shupe}, {Craig}, {Dencheva}, {Ginsburg}, {VanderPlas}, {Bradley},
  {P{\'e}rez-Su{\'a}rez}, {de Val-Borro}, {Aldcroft}, {Cruz}, {Robitaille},
  {Tollerud}, {Ardelean}, {Babej}, {Bach}, {Bachetti}, {Bakanov}, {Bamford},
  {Barentsen}, {Barmby}, {Baumbach}, {Berry}, {Biscani}, {Boquien}, {Bostroem},
  {Bouma}, {Brammer}, {Bray}, {Breytenbach}, {Buddelmeijer}, {Burke},
  {Calderone}, {Cano Rodr{\'\i}guez}, {Cara}, {Cardoso}, {Cheedella}, {Copin},
  {Corrales}, {Crichton}, {D'Avella}, {Deil}, {Depagne}, {Dietrich}, {Donath},
  {Droettboom}, {Earl}, {Erben}, {Fabbro}, {Ferreira}, {Finethy}, {Fox},
  {Garrison}, {Gibbons}, {Goldstein}, {Gommers}, {Greco}, {Greenfield},
  {Groener}, {Grollier}, {Hagen}, {Hirst}, {Homeier}, {Horton}, {Hosseinzadeh},
  {Hu}, {Hunkeler}, {Ivezi{\'c}}, {Jain}, {Jenness}, {Kanarek}, {Kendrew},
  {Kern}, {Kerzendorf}, {Khvalko}, {King}, {Kirkby}, {Kulkarni}, {Kumar},
  {Lee}, {Lenz}, {Littlefair}, {Ma}, {Macleod}, {Mastropietro}, {McCully},
  {Montagnac}, {Morris}, {Mueller}, {Mumford}, {Muna}, {Murphy}, {Nelson},
  {Nguyen}, {Ninan}, {N{\"o}the}, {Ogaz}, {Oh}, {Parejko}, {Parley}, {Pascual},
  {Patil}, {Patil}, {Plunkett}, {Prochaska}, {Rastogi}, {Reddy Janga},
  {Sabater}, {Sakurikar}, {Seifert}, {Sherbert}, {Sherwood-Taylor}, {Shih},
  {Sick}, {Silbiger}, {Singanamalla}, {Singer}, {Sladen}, {Sooley},
  {Sornarajah}, {Streicher}, {Teuben}, {Thomas}, {Tremblay}, {Turner},
  {Terr{\'o}n}, {van Kerkwijk}, {de la Vega}, {Watkins}, {Weaver}, {Whitmore},
  {Woillez}, {Zabalza}, \& {Astropy Contributors}}]{2018AJ....156..123A}
{Astropy Collaboration}, {Price-Whelan}, A.~M., {Sip{\H{o}}cz}, B.~M., {et~al.}
  2018, \aj, 156, 123, \dodoi{10.3847/1538-3881/aabc4f}

\bibitem[{{Bekenstein}(1973)}]{Bekenstein1973}
{Bekenstein}, J.~D. 1973, \apj, 183, 657, \dodoi{10.1086/152255}

\bibitem[{{Bohlin} {et~al.}(2017){Bohlin}, {M{\'e}sz{\'a}ros}, {Fleming},
  {Gordon}, {Koekemoer}, \& {Kov{\'a}cs}}]{Bohlin2017}
{Bohlin}, R.~C., {M{\'e}sz{\'a}ros}, S., {Fleming}, S.~W., {et~al.} 2017, \aj,
  153, 234, \dodoi{10.3847/1538-3881/aa6ba9}

\bibitem[{{Boisse} {et~al.}(2010){Boisse}, {Eggenberger}, {Santos}, {Lovis},
  {Bouchy}, {H{\'e}brard}, {Arnold}, {Bonfils}, {Delfosse}, {Desort},
  {D{\'\i}az}, {Ehrenreich}, {Forveille}, {Gallenne}, {Lagrange}, {Moutou},
  {Udry}, {Pepe}, {Perrier}, {Perruchot}, {Pont}, {Queloz}, {Santerne},
  {S{\'e}gransan}, \& {Vidal-Madjar}}]{Boisse2010}
{Boisse}, I., {Eggenberger}, A., {Santos}, N.~C., {et~al.} 2010, \aap, 523,
  A88, \dodoi{10.1051/0004-6361/201014909}

\bibitem[{{Brahm} {et~al.}(2017){Brahm}, {Jord{\'a}n}, \&
  {Espinoza}}]{Brahm2017}
{Brahm}, R., {Jord{\'a}n}, A., \& {Espinoza}, N. 2017, \pasp, 129, 034002,
  \dodoi{10.1088/1538-3873/aa5455}

\bibitem[{{Breivik} {et~al.}(2017){Breivik}, {Chatterjee}, \&
  {Larson}}]{breivik_2017}
{Breivik}, K., {Chatterjee}, S., \& {Larson}, S.~L. 2017, \apjl, 850, L13,
  \dodoi{10.3847/2041-8213/aa97d5}

\bibitem[{{Brown} \& {Bethe}(1994)}]{brown_bethe_1994}
{Brown}, G.~E., \& {Bethe}, H.~A. 1994, \apj, 423, 659, \dodoi{10.1086/173844}

\bibitem[{{Buchhave} {et~al.}(2010){Buchhave}, {Bakos}, {Hartman}, {Torres},
  {Kov{\'a}cs}, {Latham}, {Noyes}, {Esquerdo}, {Everett}, {Howard}, {Marcy},
  {Fischer}, {Johnson}, {Andersen}, {F{\H{u}}r{\'e}sz}, {Perumpilly},
  {Sasselov}, {Stefanik}, {B{\'e}ky}, {L{\'a}z{\'a}r}, {Papp}, \&
  {S{\'a}ri}}]{Buchhave2010}
{Buchhave}, L.~A., {Bakos}, G.~{\'A}., {Hartman}, J.~D., {et~al.} 2010, \apj,
  720, 1118, \dodoi{10.1088/0004-637X/720/2/1118}

\bibitem[{{Chakrabarti} {et~al.}(2023){Chakrabarti}, {Simon}, {Craig},
  {Reggiani}, {Brandt}, {Guhathakurta}, {Dalba}, {Kirby}, {Chang}, {Hey},
  {Savino}, {Geha}, \& {Thompson}}]{Chakrabarti2023}
{Chakrabarti}, S., {Simon}, J.~D., {Craig}, P.~A., {et~al.} 2023, \aj, 166, 6,
  \dodoi{10.3847/1538-3881/accf21}

\bibitem[{Chawla {et~al.}(2022)Chawla, Chatterjee, Breivik, Moorthy, Andrews,
  \& Sanderson}]{Chawla_2022}
Chawla, C., Chatterjee, S., Breivik, K., {et~al.} 2022, \apj, 931, 107,
  \dodoi{10.3847/1538-4357/ac60a5}

\bibitem[{{Corral-Santana} {et~al.}(2016){Corral-Santana}, {Casares},
  {Mu{\~n}oz-Darias}, {Bauer}, {Mart{\'\i}nez-Pais}, \&
  {Russell}}]{corral_santana_2016}
{Corral-Santana}, J.~M., {Casares}, J., {Mu{\~n}oz-Darias}, T., {et~al.} 2016,
  \aap, 587, A61, \dodoi{10.1051/0004-6361/201527130}

\bibitem[{{Davies} {et~al.}(2018){Davies}, {Crowther}, \&
  {Beasor}}]{davies_2018}
{Davies}, B., {Crowther}, P.~A., \& {Beasor}, E.~R. 2018, \mnras, 478, 3138,
  \dodoi{10.1093/mnras/sty1302}

\bibitem[{{Di Carlo} {et~al.}(2023){Di Carlo}, Agrawal, Rodriguez, \&
  Breivik}]{dicarlo2023young}
{Di Carlo}, U., Agrawal, P., Rodriguez, C.~L., \& Breivik, K. 2023, arXiv.
\newblock \doarXiv{2306.13121}

\bibitem[{{El-Badry} \& {Rix}(2018)}]{El-Badry2018}
{El-Badry}, K., \& {Rix}, H.-W. 2018, \mnras, 480, 4884,
  \dodoi{10.1093/mnras/sty2186}

\bibitem[{{El-Badry} {et~al.}(2023{\natexlab{a}}){El-Badry}, {Rix}, {Quataert},
  {Howard}, {Isaacson}, {Fuller}, {Hawkins}, {Breivik}, {Wong}, {Rodriguez},
  {Conroy}, {Shahaf}, {Mazeh}, {Arenou}, {Burdge}, {Bashi}, {Faigler}, {Weisz},
  {Seeburger}, {Almada Monter}, \& {Wojno}}]{Gaia_BH1}
{El-Badry}, K., {Rix}, H.-W., {Quataert}, E., {et~al.} 2023{\natexlab{a}},
  \mnras, 518, 1057, \dodoi{10.1093/mnras/stac3140}

\bibitem[{{El-Badry} {et~al.}(2023{\natexlab{b}}){El-Badry}, {Rix}, {Cendes},
  {Rodriguez}, {Conroy}, {Quataert}, {Hawkins}, {Zari}, {Hobson}, {Breivik},
  {Rau}, {Berger}, {Shahaf}, {Seeburger}, {Burdge}, {Latham}, {Buchhave},
  {Bieryla}, {Bashi}, {Mazeh}, \& {Faigler}}]{Gaia_BH2}
{El-Badry}, K., {Rix}, H.-W., {Cendes}, Y., {et~al.} 2023{\natexlab{b}},
  \mnras, 521, 4323, \dodoi{10.1093/mnras/stad799}

\bibitem[{{F{\H{u}}r{\'e}sz}(2008)}]{Furesz2008}
{F{\H{u}}r{\'e}sz}, G. 2008, PhD thesis, University of Szeged, Hungary

\bibitem[{{Foreman-Mackey} {et~al.}(2013){Foreman-Mackey}, {Hogg}, {Lang}, \&
  {Goodman}}]{emcee_2013}
{Foreman-Mackey}, D., {Hogg}, D.~W., {Lang}, D., \& {Goodman}, J. 2013, \pasp,
  125, 306, \dodoi{10.1086/670067}

\bibitem[{Fulton {et~al.}(2020)Fulton, Blunt, Hurt, Mills, cpsdoppler,
  sinukoff, Hadden, Bouma, zhexingli, iancrossfield, dos Santos, Benneke,
  Weiss, Yee, Rosenthal, Hirsch, Petigura, \& Foreman-Mackey}]{radvel_zenodo}
Fulton, B., Blunt, S., Hurt, S., {et~al.} 2020,
  California-Planet-Search/radvel: Version 1.4.0, v1.4.0,  Zenodo,
  \dodoi{10.5281/zenodo.3814717}

\bibitem[{{Fulton} {et~al.}(2018){Fulton}, {Petigura}, {Blunt}, \&
  {Sinukoff}}]{fulton_petigura_blunt_sinukoff_2018}
{Fulton}, B.~J., {Petigura}, E.~A., {Blunt}, S., \& {Sinukoff}, E. 2018, \pasp,
  130, 044504, \dodoi{10.1088/1538-3873/aaaaa8}

\bibitem[{{Gaia Collaboration} {et~al.}(2023{\natexlab{a}}){Gaia
  Collaboration}, {Vallenari}, {Brown}, {Prusti}, {de Bruijne}, {Arenou},
  {Babusiaux}, {Biermann}, {Creevey}, {Ducourant}, {Evans}, {Eyer}, {Guerra},
  {Hutton}, {Jordi}, {Klioner}, {Lammers}, {Lindegren}, {Luri}, {Mignard},
  {Panem}, {Pourbaix}, {Randich}, {Sartoretti}, {Soubiran}, {Tanga}, {Walton},
  {Bailer-Jones}, {Bastian}, {Drimmel}, {Jansen}, {Katz}, {Lattanzi}, {van
  Leeuwen}, {Bakker}, {Cacciari}, {Casta{\~n}eda}, {De Angeli}, {Fabricius},
  {Fouesneau}, {Fr{\'e}mat}, {Galluccio}, {Guerrier}, {Heiter}, {Masana},
  {Messineo}, {Mowlavi}, {Nicolas}, {Nienartowicz}, {Pailler}, {Panuzzo},
  {Riclet}, {Roux}, {Seabroke}, {Sordo}, {Th{\'e}venin}, {Gracia-Abril},
  {Portell}, {Teyssier}, {Altmann}, {Andrae}, {Audard}, {Bellas-Velidis},
  {Benson}, {Berthier}, {Blomme}, {Burgess}, {Busonero}, {Busso},
  {C{\'a}novas}, {Carry}, {Cellino}, {Cheek}, {Clementini}, {Damerdji},
  {Davidson}, {de Teodoro}, {Nu{\~n}ez Campos}, {Delchambre}, {Dell'Oro},
  {Esquej}, {Fern{\'a}ndez-Hern{\'a}ndez}, {Fraile}, {Garabato},
  {Garc{\'\i}a-Lario}, {Gosset}, {Haigron}, {Halbwachs}, {Hambly}, {Harrison},
  {Hern{\'a}ndez}, {Hestroffer}, {Hodgkin}, {Holl}, {Jan{\ss}en}, {Jevardat de
  Fombelle}, {Jordan}, {Krone-Martins}, {Lanzafame}, {L{\"o}ffler}, {Marchal},
  {Marrese}, {Moitinho}, {Muinonen}, {Osborne}, {Pancino}, {Pauwels},
  {Recio-Blanco}, {Reyl{\'e}}, {Riello}, {Rimoldini}, {Roegiers}, {Rybizki},
  {Sarro}, {Siopis}, {Smith}, {Sozzetti}, {Utrilla}, {van Leeuwen}, {Abbas},
  {{\'A}brah{\'a}m}, {Abreu Aramburu}, {Aerts}, {Aguado}, {Ajaj},
  {Aldea-Montero}, {Altavilla}, {{\'A}lvarez}, {Alves}, {Anders}, {Anderson},
  {Anglada Varela}, {Antoja}, {Baines}, {Baker}, {Balaguer-N{\'u}{\~n}ez},
  {Balbinot}, {Balog}, {Barache}, {Barbato}, {Barros}, {Barstow},
  {Bartolom{\'e}}, {Bassilana}, {Bauchet}, {Becciani}, {Bellazzini},
  {Berihuete}, {Bernet}, {Bertone}, {Bianchi}, {Binnenfeld}, {Blanco-Cuaresma},
  {Blazere}, {Boch}, {Bombrun}, {Bossini}, {Bouquillon}, {Bragaglia},
  {Bramante}, {Breedt}, {Bressan}, {Brouillet}, {Brugaletta}, {Bucciarelli},
  {Burlacu}, {Butkevich}, {Buzzi}, {Caffau}, {Cancelliere}, {Cantat-Gaudin},
  {Carballo}, {Carlucci}, {Carnerero}, {Carrasco}, {Casamiquela}, {Castellani},
  {Castro-Ginard}, {Chaoul}, {Charlot}, {Chemin}, {Chiaramida}, {Chiavassa},
  {Chornay}, {Comoretto}, {Contursi}, {Cooper}, {Cornez}, {Cowell}, {Crifo},
  {Cropper}, {Crosta}, {Crowley}, {Dafonte}, {Dapergolas}, {David}, {David},
  {de Laverny}, {De Luise}, {De March}, {De Ridder}, {de Souza}, {de Torres},
  {del Peloso}, {del Pozo}, {Delbo}, {Delgado}, {Delisle}, {Demouchy},
  {Dharmawardena}, {Di Matteo}, {Diakite}, {Diener}, {Distefano}, {Dolding},
  {Edvardsson}, {Enke}, {Fabre}, {Fabrizio}, {Faigler}, {Fedorets}, {Fernique},
  {Fienga}, {Figueras}, {Fournier}, {Fouron}, {Fragkoudi}, {Gai},
  {Garcia-Gutierrez}, {Garcia-Reinaldos}, {Garc{\'\i}a-Torres}, {Garofalo},
  {Gavel}, {Gavras}, {Gerlach}, {Geyer}, {Giacobbe}, {Gilmore}, {Girona},
  {Giuffrida}, {Gomel}, {Gomez}, {Gonz{\'a}lez-N{\'u}{\~n}ez},
  {Gonz{\'a}lez-Santamar{\'\i}a}, {Gonz{\'a}lez-Vidal}, {Granvik}, {Guillout},
  {Guiraud}, {Guti{\'e}rrez-S{\'a}nchez}, {Guy}, {Hatzidimitriou}, {Hauser},
  {Haywood}, {Helmer}, {Helmi}, {Sarmiento}, {Hidalgo}, {Hilger},
  {H{\l}adczuk}, {Hobbs}, {Holland}, {Huckle}, {Jardine}, {Jasniewicz},
  {Jean-Antoine Piccolo}, {Jim{\'e}nez-Arranz}, {Jorissen}, {Juaristi
  Campillo}, {Julbe}, {Karbevska}, {Kervella}, {Khanna}, {Kontizas},
  {Kordopatis}, {Korn}, {K{\'o}sp{\'a}l}, {Kostrzewa-Rutkowska},
  {Kruszy{\'n}ska}, {Kun}, {Laizeau}, {Lambert}, {Lanza}, {Lasne}, {Le
  Campion}, {Lebreton}, {Lebzelter}, {Leccia}, {Leclerc}, {Lecoeur-Taibi},
  {Liao}, {Licata}, {Lindstr{\o}m}, {Lister}, {Livanou}, {Lobel}, {Lorca},
  {Loup}, {Madrero Pardo}, {Magdaleno Romeo}, {Managau}, {Mann}, {Manteiga},
  {Marchant}, {Marconi}, {Marcos}, {Marcos Santos}, {Mar{\'\i}n Pina},
  {Marinoni}, {Marocco}, {Marshall}, {Martin Polo}, {Mart{\'\i}n-Fleitas},
  {Marton}, {Mary}, {Masip}, {Massari}, {Mastrobuono-Battisti}, {Mazeh},
  {McMillan}, {Messina}, {Michalik}, {Millar}, {Mints}, {Molina}, {Molinaro},
  {Moln{\'a}r}, {Monari}, {Mongui{\'o}}, {Montegriffo}, {Montero}, {Mor},
  {Mora}, {Morbidelli}, {Morel}, {Morris}, {Muraveva}, {Murphy}, {Musella},
  {Nagy}, {Noval}, {Oca{\~n}a}, {Ogden}, {Ordenovic}, {Osinde}, {Pagani},
  {Pagano}, {Palaversa}, {Palicio}, {Pallas-Quintela}, {Panahi},
  {Payne-Wardenaar}, {Pe{\~n}alosa Esteller}, {Penttil{\"a}}, {Pichon},
  {Piersimoni}, {Pineau}, {Plachy}, {Plum}, {Poggio}, {Pr{\v{s}}a}, {Pulone},
  {Racero}, {Ragaini}, {Rainer}, {Raiteri}, {Rambaux}, {Ramos}, {Ramos-Lerate},
  {Re Fiorentin}, {Regibo}, {Richards}, {Rios Diaz}, {Ripepi}, {Riva}, {Rix},
  {Rixon}, {Robichon}, {Robin}, {Robin}, {Roelens}, {Rogues}, {Rohrbasser},
  {Romero-G{\'o}mez}, {Rowell}, {Royer}, {Ruz Mieres}, {Rybicki}, {Sadowski},
  {S{\'a}ez N{\'u}{\~n}ez}, {Sagrist{\`a} Sell{\'e}s}, {Sahlmann}, {Salguero},
  {Samaras}, {Sanchez Gimenez}, {Sanna}, {Santove{\~n}a}, {Sarasso},
  {Schultheis}, {Sciacca}, {Segol}, {Segovia}, {S{\'e}gransan}, {Semeux},
  {Shahaf}, {Siddiqui}, {Siebert}, {Siltala}, {Silvelo}, {Slezak}, {Slezak},
  {Smart}, {Snaith}, {Solano}, {Solitro}, {Souami}, {Souchay}, {Spagna},
  {Spina}, {Spoto}, {Steele}, {Steidelm{\"u}ller}, {Stephenson}, {S{\"u}veges},
  {Surdej}, {Szabados}, {Szegedi-Elek}, {Taris}, {Taylor}, {Teixeira},
  {Tolomei}, {Tonello}, {Torra}, {Torra}, {Torralba Elipe}, {Trabucchi},
  {Tsounis}, {Turon}, {Ulla}, {Unger}, {Vaillant}, {van Dillen}, {van Reeven},
  {Vanel}, {Vecchiato}, {Viala}, {Vicente}, {Voutsinas}, {Weiler}, {Wevers},
  {Wyrzykowski}, {Yoldas}, {Yvard}, {Zhao}, {Zorec}, {Zucker}, \&
  {Zwitter}}]{gaia_dr3_summary_2023}
{Gaia Collaboration}, {Vallenari}, A., {Brown}, A.~G.~A., {et~al.}
  2023{\natexlab{a}}, \aap, 674, A1, \dodoi{10.1051/0004-6361/202243940}

\bibitem[{{Gaia Collaboration} {et~al.}(2023{\natexlab{b}}){Gaia
  Collaboration}, {Arenou}, {Babusiaux}, {Barstow}, {Faigler}, {Jorissen},
  {Kervella}, {Mazeh}, {Mowlavi}, {Panuzzo}, {Sahlmann}, {Shahaf}, {Sozzetti},
  {Bauchet}, {Damerdji}, {Gavras}, {Giacobbe}, {Gosset}, {Halbwachs}, {Holl},
  {Lattanzi}, {Leclerc}, {Morel}, {Pourbaix}, {Re Fiorentin}, {Sadowski},
  {S{\'e}gransan}, {Siopis}, {Teyssier}, {Zwitter}, {Planquart}, {Brown},
  {Vallenari}, {Prusti}, {de Bruijne}, {Biermann}, {Creevey}, {Ducourant},
  {Evans}, {Eyer}, {Guerra}, {Hutton}, {Jordi}, {Klioner}, {Lammers},
  {Lindegren}, {Luri}, {Mignard}, {Panem}, {Randich}, {Sartoretti}, {Soubiran},
  {Tanga}, {Walton}, {Bailer-Jones}, {Bastian}, {Drimmel}, {Jansen}, {Katz},
  {van Leeuwen}, {Bakker}, {Cacciari}, {Casta{\~n}eda}, {De Angeli},
  {Fabricius}, {Fouesneau}, {Fr{\'e}mat}, {Galluccio}, {Guerrier}, {Heiter},
  {Masana}, {Messineo}, {Nicolas}, {Nienartowicz}, {Pailler}, {Riclet}, {Roux},
  {Seabroke}, {Sordo}, {Th{\'e}venin}, {Gracia-Abril}, {Portell}, {Altmann},
  {Andrae}, {Audard}, {Bellas-Velidis}, {Benson}, {Berthier}, {Blomme},
  {Burgess}, {Busonero}, {Busso}, {C{\'a}novas}, {Carry}, {Cellino}, {Cheek},
  {Clementini}, {Davidson}, {de Teodoro}, {Nu{\~n}ez Campos}, {Delchambre},
  {Dell'Oro}, {Esquej}, {Fern{\'a}ndez-Hern{\'a}ndez}, {Fraile}, {Garabato},
  {Garc{\'\i}a-Lario}, {Haigron}, {Hambly}, {Harrison}, {Hern{\'a}ndez},
  {Hestroffer}, {Hodgkin}, {Jan{\ss}en}, {Jevardat de Fombelle}, {Jordan},
  {Krone-Martins}, {Lanzafame}, {L{\"o}ffler}, {Marchal}, {Marrese},
  {Moitinho}, {Muinonen}, {Osborne}, {Pancino}, {Pauwels}, {Recio-Blanco},
  {Reyl{\'e}}, {Riello}, {Rimoldini}, {Roegiers}, {Rybizki}, {Sarro}, {Smith},
  {Utrilla}, {van Leeuwen}, {Abbas}, {{\'A}brah{\'a}m}, {Abreu Aramburu},
  {Aerts}, {Aguado}, {Ajaj}, {Aldea-Montero}, {Altavilla}, {{\'A}lvarez},
  {Alves}, {Anders}, {Anderson}, {Anglada Varela}, {Antoja}, {Baines}, {Baker},
  {Balaguer-N{\'u}{\~n}ez}, {Balbinot}, {Balog}, {Barache}, {Barbato},
  {Barros}, {Bartolom{\'e}}, {Bassilana}, {Becciani}, {Bellazzini},
  {Berihuete}, {Bernet}, {Bertone}, {Bianchi}, {Binnenfeld}, {Blanco-Cuaresma},
  {Blazere}, {Boch}, {Bombrun}, {Bossini}, {Bouquillon}, {Bragaglia},
  {Bramante}, {Breedt}, {Bressan}, {Brouillet}, {Brugaletta}, {Bucciarelli},
  {Burlacu}, {Butkevich}, {Buzzi}, {Caffau}, {Cancelliere}, {Cantat-Gaudin},
  {Carballo}, {Carlucci}, {Carnerero}, {Carrasco}, {Casamiquela}, {Castellani},
  {Castro-Ginard}, {Chaoul}, {Charlot}, {Chemin}, {Chiaramida}, {Chiavassa},
  {Chornay}, {Comoretto}, {Contursi}, {Cooper}, {Cornez}, {Cowell}, {Crifo},
  {Cropper}, {Crosta}, {Crowley}, {Dafonte}, {Dapergolas}, {David}, {de
  Laverny}, {De Luise}, {De March}, {De Ridder}, {de Souza}, {de Torres}, {del
  Peloso}, {del Pozo}, {Delbo}, {Delgado}, {Delisle}, {Demouchy},
  {Dharmawardena}, {Diakite}, {Diener}, {Distefano}, {Dolding}, {Enke},
  {Fabre}, {Fabrizio}, {Fedorets}, {Fernique}, {Figueras}, {Fournier},
  {Fouron}, {Fragkoudi}, {Gai}, {Garcia-Gutierrez}, {Garcia-Reinaldos},
  {Garc{\'\i}a-Torres}, {Garofalo}, {Gavel}, {Gerlach}, {Geyer}, {Gilmore},
  {Girona}, {Giuffrida}, {Gomel}, {Gomez}, {Gonz{\'a}lez-N{\'u}{\~n}ez},
  {Gonz{\'a}lez-Santamar{\'\i}a}, {Gonz{\'a}lez-Vidal}, {Granvik}, {Guillout},
  {Guiraud}, {Guti{\'e}rrez-S{\'a}nchez}, {Guy}, {Hatzidimitriou}, {Hauser},
  {Haywood}, {Helmer}, {Helmi}, {Sarmiento}, {Hidalgo}, {Hilger},
  {H{\l}adczuk}, {Hobbs}, {Holland}, {Huckle}, {Jardine}, {Jasniewicz},
  {Jean-Antoine Piccolo}, {Jim{\'e}nez-Arranz}, {Juaristi Campillo}, {Julbe},
  {Karbevska}, {Khanna}, {Kordopatis}, {Korn}, {K{\'o}sp{\'a}l},
  {Kostrzewa-Rutkowska}, {Kruszy{\'n}ska}, {Kun}, {Laizeau}, {Lambert},
  {Lanza}, {Lasne}, {Le Campion}, {Lebreton}, {Lebzelter}, {Leccia},
  {Lecoeur-Taibi}, {Liao}, {Licata}, {Lindstr{\o}m}, {Lister}, {Livanou},
  {Lobel}, {Lorca}, {Loup}, {Madrero Pardo}, {Magdaleno Romeo}, {Managau},
  {Mann}, {Manteiga}, {Marchant}, {Marconi}, {Marcos}, {Marcos Santos},
  {Mar{\'\i}n Pina}, {Marinoni}, {Marocco}, {Marshall}, {Martin Polo},
  {Mart{\'\i}n-Fleitas}, {Marton}, {Mary}, {Masip}, {Massari},
  {Mastrobuono-Battisti}, {McMillan}, {Messina}, {Michalik}, {Millar}, {Mints},
  {Molina}, {Molinaro}, {Moln{\'a}r}, {Monari}, {Mongui{\'o}}, {Montegriffo},
  {Montero}, {Mor}, {Mora}, {Morbidelli}, {Morris}, {Muraveva}, {Murphy},
  {Musella}, {Nagy}, {Noval}, {Oca{\~n}a}, {Ogden}, {Ordenovic}, {Osinde},
  {Pagani}, {Pagano}, {Palaversa}, {Palicio}, {Pallas-Quintela}, {Panahi},
  {Payne-Wardenaar}, {Pe{\~n}alosa Esteller}, {Penttil{\"a}}, {Pichon},
  {Piersimoni}, {Pineau}, {Plachy}, {Plum}, {Poggio}, {Pr{\v{s}}a}, {Pulone},
  {Racero}, {Ragaini}, {Rainer}, {Raiteri}, {Ramos}, {Ramos-Lerate}, {Regibo},
  {Richards}, {Rios Diaz}, {Ripepi}, {Riva}, {Rix}, {Rixon}, {Robichon},
  {Robin}, {Robin}, {Roelens}, {Rogues}, {Rohrbasser}, {Romero-G{\'o}mez},
  {Rowell}, {Royer}, {Ruz Mieres}, {Rybicki}, {S{\'a}ez N{\'u}{\~n}ez},
  {Sagrist{\`a} Sell{\'e}s}, {Salguero}, {Samaras}, {Sanchez Gimenez}, {Sanna},
  {Santove{\~n}a}, {Sarasso}, {Schultheis}, {Sciacca}, {Segol}, {Segovia},
  {Semeux}, {Siddiqui}, {Siebert}, {Siltala}, {Silvelo}, {Slezak}, {Slezak},
  {Smart}, {Snaith}, {Solano}, {Solitro}, {Souami}, {Souchay}, {Spagna},
  {Spina}, {Spoto}, {Steele}, {Steidelm{\"u}ller}, {Stephenson}, {S{\"u}veges},
  {Surdej}, {Szabados}, {Szegedi-Elek}, {Taris}, {Taylor}, {Teixeira},
  {Tolomei}, {Tonello}, {Torra}, {Torra}, {Torralba Elipe}, {Trabucchi},
  {Tsounis}, {Turon}, {Ulla}, {Unger}, {Vaillant}, {van Dillen}, {van Reeven},
  {Vanel}, {Vecchiato}, {Viala}, {Vicente}, {Voutsinas}, {Weiler}, {Wevers},
  {Wyrzykowski}, {Yoldas}, {Yvard}, {Zhao}, {Zorec}, \&
  {Zucker}}]{gaia_dr3_binaries_2023}
{Gaia Collaboration}, {Arenou}, F., {Babusiaux}, C., {et~al.}
  2023{\natexlab{b}}, \aap, 674, A34, \dodoi{10.1051/0004-6361/202243782}

\bibitem[{{Gibson} {et~al.}(2016){Gibson}, {Howard}, {Marcy}, {Edelstein},
  {Wishnow}, \& {Poppett}}]{Gibson2016}
{Gibson}, S.~R., {Howard}, A.~W., {Marcy}, G.~W., {et~al.} 2016, in Society of
  Photo-Optical Instrumentation Engineers (SPIE) Conference Series, Vol. 9908,
  Ground-based and Airborne Instrumentation for Astronomy VI, ed. C.~J.
  {Evans}, L.~{Simard}, \& H.~{Takami}, 990870, \dodoi{10.1117/12.2233334}

\bibitem[{{Giesers} {et~al.}(2018){Giesers}, {Dreizler}, {Husser}, {Kamann},
  {Anglada Escud{\'e}}, {Brinchmann}, {Carollo}, {Roth}, {Weilbacher}, \&
  {Wisotzki}}]{giesers_2018}
{Giesers}, B., {Dreizler}, S., {Husser}, T.-O., {et~al.} 2018, \mnras, 475,
  L15, \dodoi{10.1093/mnrasl/slx203}

\bibitem[{{Gilkis} {et~al.}(2021){Gilkis}, {Shenar}, {Ramachandran}, {Jermyn},
  {Mahy}, {Oskinova}, {Arcavi}, \& {Sana}}]{gilkis_2021}
{Gilkis}, A., {Shenar}, T., {Ramachandran}, V., {et~al.} 2021, \mnras, 503,
  1884, \dodoi{10.1093/mnras/stab383}

\bibitem[{{Gray}(2008)}]{Gray2008}
{Gray}, D.~F. 2008, {The Observation and Analysis of Stellar Photospheres}

\bibitem[{{Guseinov} \& {Zel'dovich}(1966)}]{guseinov_1966}
{Guseinov}, O.~K., \& {Zel'dovich}, Y.~B. 1966, \sovast, 10, 251

\bibitem[{{Halbwachs} {et~al.}(2023){Halbwachs}, {Pourbaix}, {Arenou},
  {Galluccio}, {Guillout}, {Bauchet}, {Marchal}, {Sadowski}, \&
  {Teyssier}}]{halbwachs_2023}
{Halbwachs}, J.-L., {Pourbaix}, D., {Arenou}, F., {et~al.} 2023, \aap, 674, A9,
  \dodoi{10.1051/0004-6361/202243969}

\bibitem[{{Hayashi} \& {Suto}(2020)}]{hayashi_suto_2020}
{Hayashi}, T., \& {Suto}, Y. 2020, \apj, 897, 29,
  \dodoi{10.3847/1538-4357/ab97ad}

\bibitem[{{Hayashi} {et~al.}(2023){Hayashi}, {Suto}, \&
  {Trani}}]{hayashi_suto_trani_2023}
{Hayashi}, T., {Suto}, Y., \& {Trani}, A.~A. 2023, \apj, 958, 26,
  \dodoi{10.3847/1538-4357/acf4f6}

\bibitem[{{Hayashi} {et~al.}(2020){Hayashi}, {Wang}, \&
  {Suto}}]{hayashi_wang_suto_2020}
{Hayashi}, T., {Wang}, S., \& {Suto}, Y. 2020, \apj, 890, 112,
  \dodoi{10.3847/1538-4357/ab6de6}

\bibitem[{{Hirai} \& {Mandel}(2022)}]{Hirai2022}
{Hirai}, R., \& {Mandel}, I. 2022, \apjl, 937, L42,
  \dodoi{10.3847/2041-8213/ac9519}

\bibitem[{{Howell} \& {Furlan}(2022)}]{2022FrASS...9.1163H}
{Howell}, S.~B., \& {Furlan}, E. 2022, Frontiers in Astronomy and Space
  Sciences, 9, 871163, \dodoi{10.3389/fspas.2022.871163}

\bibitem[{{Humphreys} \& {Davidson}(1979)}]{HD_1979}
{Humphreys}, R.~M., \& {Davidson}, K. 1979, \apj, 232, 409,
  \dodoi{10.1086/157301}

\bibitem[{{Janssens} {et~al.}(2022){Janssens}, {Shenar}, {Sana}, {Faigler},
  {Langer}, {Marchant}, {Mazeh}, {Sch{\"u}rmann}, \& {Shahaf}}]{janssens_2022}
{Janssens}, S., {Shenar}, T., {Sana}, H., {et~al.} 2022, \aap, 658, A129,
  \dodoi{10.1051/0004-6361/202141866}

\bibitem[{{Justham} {et~al.}(2014){Justham}, {Podsiadlowski}, \&
  {Vink}}]{Justham2014}
{Justham}, S., {Podsiadlowski}, P., \& {Vink}, J.~S. 2014, \apj, 796, 121,
  \dodoi{10.1088/0004-637X/796/2/121}

\bibitem[{{Kaufer} {et~al.}(1999){Kaufer}, {Stahl}, {Tubbesing},
  {N{\o}rregaard}, {Avila}, {Francois}, {Pasquini}, \& {Pizzella}}]{Kaufer1999}
{Kaufer}, A., {Stahl}, O., {Tubbesing}, S., {et~al.} 1999, The Messenger, 95, 8

\bibitem[{Koposov {et~al.}(2023)Koposov, Speagle, Barbary, Ashton, Bennett,
  Buchner, Scheffler, Cook, Talbot, Guillochon, Cubillos, Ramos, Johnson, Lang,
  Ilya, Dartiailh, Nitz, McCluskey, \& Archibald}]{koposov_2023}
Koposov, S., Speagle, J., Barbary, K., {et~al.} 2023, joshspeagle/dynesty:
  v2.1.3, v2.1.3,  Zenodo, \dodoi{10.5281/zenodo.8408702}

\bibitem[{{Kozai}(1962)}]{1962AJ.....67..591K}
{Kozai}, Y. 1962, \aj, 67, 591, \dodoi{10.1086/108790}

\bibitem[{{Kurucz}(1979)}]{Kurucz1979}
{Kurucz}, R.~L. 1979, \apjs, 40, 1, \dodoi{10.1086/190589}

\bibitem[{{Kurucz}(1993)}]{Kurucz1993}
---. 1993, {SYNTHE spectrum synthesis programs and line data}

\bibitem[{{Lafarga} {et~al.}(2020){Lafarga}, {Ribas}, {Lovis}, {Perger},
  {Zechmeister}, {Bauer}, {K{\"u}rster}, {Cort{\'e}s-Contreras}, {Morales},
  {Herrero}, {Rosich}, {Baroch}, {Reiners}, {Caballero}, {Quirrenbach},
  {Amado}, {Alacid}, {B{\'e}jar}, {Dreizler}, {Hatzes}, {Henning}, {Jeffers},
  {Kaminski}, {Montes}, {Pedraz}, {Rodr{\'\i}guez-L{\'o}pez}, \&
  {Schmitt}}]{Lafarga2020}
{Lafarga}, M., {Ribas}, I., {Lovis}, C., {et~al.} 2020, \aap, 636, A36,
  \dodoi{10.1051/0004-6361/201937222}

\bibitem[{{Lidov}(1962)}]{1962P&SS....9..719L}
{Lidov}, M.~L. 1962, \planss, 9, 719, \dodoi{10.1016/0032-0633(62)90129-0}

\bibitem[{{Lindegren} \& {Dravins}(2003)}]{lindegren_2003}
{Lindegren}, L., \& {Dravins}, D. 2003, \aap, 401, 1185,
  \dodoi{10.1051/0004-6361:20030181}

\bibitem[{{Liu} {et~al.}(2022){Liu}, {D'Orazio}, {Vigna-G{\'o}mez}, \&
  {Samsing}}]{Liu2022}
{Liu}, B., {D'Orazio}, D.~J., {Vigna-G{\'o}mez}, A., \& {Samsing}, J. 2022,
  \prd, 106, 123010, \dodoi{10.1103/PhysRevD.106.123010}

\bibitem[{{Luhn} {et~al.}(2020){Luhn}, {Wright}, {Howard}, \&
  {Isaacson}}]{luhn_2020}
{Luhn}, J.~K., {Wright}, J.~T., {Howard}, A.~W., \& {Isaacson}, H. 2020, \aj,
  159, 235, \dodoi{10.3847/1538-3881/ab855a}

\bibitem[{{Merritt} {et~al.}(2004){Merritt}, {Milosavljevi{\'c}}, {Favata},
  {Hughes}, \& {Holz}}]{Merritt2004}
{Merritt}, D., {Milosavljevi{\'c}}, M., {Favata}, M., {Hughes}, S.~A., \&
  {Holz}, D.~E. 2004, \apjl, 607, L9, \dodoi{10.1086/421551}

\bibitem[{{Moe} \& {Di Stefano}(2017)}]{Moe2017}
{Moe}, M., \& {Di Stefano}, R. 2017, \apjs, 230, 15,
  \dodoi{10.3847/1538-4365/aa6fb6}

\bibitem[{{Morais} \& {Correia}(2008)}]{morais_correia_2008}
{Morais}, M.~H.~M., \& {Correia}, A.~C.~M. 2008, \aap, 491, 899,
  \dodoi{10.1051/0004-6361:200810741}

\bibitem[{{Morais} \& {Correia}(2011)}]{morais_correia_2011}
---. 2011, \aap, 525, A152, \dodoi{10.1051/0004-6361/201014812}

\bibitem[{{Naoz}(2016)}]{naoz_review_2016}
{Naoz}, S. 2016, \araa, 54, 441, \dodoi{10.1146/annurev-astro-081915-023315}

\bibitem[{{Pepe} {et~al.}(2002){Pepe}, {Mayor}, {Galland}, {Naef}, {Queloz},
  {Santos}, {Udry}, \& {Burnet}}]{Pepe2002}
{Pepe}, F., {Mayor}, M., {Galland}, F., {et~al.} 2002, \aap, 388, 632,
  \dodoi{10.1051/0004-6361:20020433}

\bibitem[{{Pepe} {et~al.}(2021){Pepe}, {Cristiani}, {Rebolo}, {Santos},
  {Dekker}, {Cabral}, {Di Marcantonio}, {Figueira}, {Lo Curto}, {Lovis},
  {Mayor}, {M{\'e}gevand}, {Molaro}, {Riva}, {Zapatero Osorio}, {Amate},
  {Manescau}, {Pasquini}, {Zerbi}, {Adibekyan}, {Abreu}, {Affolter}, {Alibert},
  {Aliverti}, {Allart}, {Allende Prieto}, {{\'A}lvarez}, {Alves}, {Avila},
  {Baldini}, {Bandy}, {Barros}, {Benz}, {Bianco}, {Borsa}, {Bourrier},
  {Bouchy}, {Broeg}, {Calderone}, {Cirami}, {Coelho}, {Conconi}, {Coretti},
  {Cumani}, {Cupani}, {D'Odorico}, {Damasso}, {Deiries}, {Delabre},
  {Demangeon}, {Dumusque}, {Ehrenreich}, {Faria}, {Fragoso}, {Genolet},
  {Genoni}, {G{\'e}nova Santos}, {Gonz{\'a}lez Hern{\'a}ndez}, {Hughes},
  {Iwert}, {Kerber}, {Knudstrup}, {Landoni}, {Lavie}, {Lillo-Box}, {Lizon},
  {Maire}, {Martins}, {Mehner}, {Micela}, {Modigliani}, {Monteiro}, {Monteiro},
  {Moschetti}, {Murphy}, {Nunes}, {Oggioni}, {Oliveira}, {Oshagh}, {Pall{\'e}},
  {Pariani}, {Poretti}, {Rasilla}, {Rebord{\~a}o}, {Redaelli}, {Santana
  Tschudi}, {Santin}, {Santos}, {S{\'e}gransan}, {Schmidt}, {Segovia},
  {Sosnowska}, {Sozzetti}, {Sousa}, {Span{\`o}}, {Su{\'a}rez Mascare{\~n}o},
  {Tabernero}, {Tenegi}, {Udry}, \& {Zanutta}}]{Pepe2021}
{Pepe}, F., {Cristiani}, S., {Rebolo}, R., {et~al.} 2021, \aap, 645, A96,
  \dodoi{10.1051/0004-6361/202038306}

\bibitem[{Peters(1964)}]{PhysRev.136.B1224}
Peters, P.~C. 1964, Phys. Rev., 136, B1224, \dodoi{10.1103/PhysRev.136.B1224}

\bibitem[{{Portegies Zwart} {et~al.}(1997){Portegies Zwart}, {Verbunt}, \&
  {Ergma}}]{portegeis_zwart_1997}
{Portegies Zwart}, S.~F., {Verbunt}, F., \& {Ergma}, E. 1997, \aap, 321, 207,
  \dodoi{10.48550/arXiv.astro-ph/9701037}

\bibitem[{{Price-Whelan} {et~al.}(2017){Price-Whelan}, {Hogg},
  {Foreman-Mackey}, \& {Rix}}]{Price-Whelan2017}
{Price-Whelan}, A.~M., {Hogg}, D.~W., {Foreman-Mackey}, D., \& {Rix}, H.-W.
  2017, \apj, 837, 20, \dodoi{10.3847/1538-4357/aa5e50}

\bibitem[{Rastello {et~al.}(2023)Rastello, Iorio, Mapelli, Arca-Sedda, Carlo,
  Escobar, Torniamenti, \& Shenar}]{rastello2023dynamical}
Rastello, S., Iorio, G., Mapelli, M., {et~al.} 2023, arXiv.
\newblock \doarXiv{2306.14679}

\bibitem[{{Rein} \& {Liu}(2012)}]{rebound_2012}
{Rein}, H., \& {Liu}, S.~F. 2012, \aap, 537, A128,
  \dodoi{10.1051/0004-6361/201118085}

\bibitem[{Rein {et~al.}(2023)Rein, Tamayo, Liu, Winkler, Bartram, Silburt,
  Baronett, Brown, Melikyan, Dorsey, Biscani, Rieder, Spiegel, Zhang, Pham,
  katvolk, Castro, amarillons, adalava, Foreman-Mackey, Stephen, Schurov,
  Guillochon, calixte07, Aye, jsuchecki, Conroy, Meldonization, LeBlanc, \&
  Ruth-Huang6012}]{rebound_zenodo}
Rein, H., Tamayo, D., Liu, S., {et~al.} 2023, hannorein/rebound: 4.0.1, 4.0.1,
  Zenodo, \dodoi{10.5281/zenodo.10121021}

\bibitem[{Remillard \& McClintock(2006)}]{remillard_mcclintock_2006}
Remillard, R.~A., \& McClintock, J.~E. 2006, \araa, 44, 49,
  \dodoi{10.1146/annurev.astro.44.051905.092532}

\bibitem[{Schwarz(1978)}]{schwarz}
Schwarz, G. 1978, The Annals of Statistics, 6, 461 ,
  \dodoi{10.1214/aos/1176344136}

\bibitem[{{Shahaf} {et~al.}(2023){Shahaf}, {Hallakoun}, {Mazeh}, {Ben-Ami},
  {Rekhi}, {El-Badry}, \& {Toonen}}]{Shahaf2023}
{Shahaf}, S., {Hallakoun}, N., {Mazeh}, T., {et~al.} 2023, arXiv e-prints,
  arXiv:2309.15143, \dodoi{10.48550/arXiv.2309.15143}

\bibitem[{{Shao} \& {Li}(2021)}]{Shao2021}
{Shao}, Y., \& {Li}, X.-D. 2021, \apj, 920, 81,
  \dodoi{10.3847/1538-4357/ac173e}

\bibitem[{{Shenar} {et~al.}(2022){Shenar}, {Sana}, {Mahy}, {El-Badry},
  {Marchant}, {Langer}, {Hawcroft}, {Fabry}, {Sen}, {Almeida}, {Abdul-Masih},
  {Bodensteiner}, {Crowther}, {Gieles}, {Gromadzki}, {H{\'e}nault-Brunet},
  {Herrero}, {de Koter}, {Iwanek}, {Koz{\l}owski}, {Lennon}, {Ma{\'\i}z
  Apell{\'a}niz}, {Mr{\'o}z}, {Moffat}, {Picco}, {Pietrukowicz}, {Poleski},
  {Rybicki}, {Schneider}, {Skowron}, {Skowron}, {Soszy{\'n}ski},
  {Szyma{\'n}ski}, {Toonen}, {Udalski}, {Ulaczyk}, {Vink}, \&
  {Wrona}}]{shenar_2022}
{Shenar}, T., {Sana}, H., {Mahy}, L., {et~al.} 2022, Nature Astronomy, 6, 1085,
  \dodoi{10.1038/s41550-022-01730-y}

\bibitem[{{Silsbee} \& {Tremaine}(2017)}]{Silsbee2017}
{Silsbee}, K., \& {Tremaine}, S. 2017, \apj, 836, 39,
  \dodoi{10.3847/1538-4357/aa5729}

\bibitem[{{Speagle}(2020)}]{dynesty_2020}
{Speagle}, J.~S. 2020, \mnras, 493, 3132, \dodoi{10.1093/mnras/staa278}

\bibitem[{{Tanikawa} {et~al.}(2023){Tanikawa}, {Cary}, {Shikauchi}, {Wang}, \&
  {Fujii}}]{Tanikawa2023}
{Tanikawa}, A., {Cary}, S., {Shikauchi}, M., {Wang}, L., \& {Fujii}, M.~S.
  2023, arXiv e-prints, arXiv:2303.05743, \dodoi{10.48550/arXiv.2303.05743}

\bibitem[{{Toonen} {et~al.}(2020){Toonen}, {Portegies Zwart}, {Hamers}, \&
  {Bandopadhyay}}]{Toonen2020}
{Toonen}, S., {Portegies Zwart}, S., {Hamers}, A.~S., \& {Bandopadhyay}, D.
  2020, \aap, 640, A16, \dodoi{10.1051/0004-6361/201936835}

\bibitem[{{Trimble} \& {Thorne}(1969)}]{trimble_thorne_1969}
{Trimble}, V.~L., \& {Thorne}, K.~S. 1969, \apj, 156, 1013,
  \dodoi{10.1086/150032}

\bibitem[{{Vogt} {et~al.}(1994){Vogt}, {Allen}, {Bigelow}, {Bresee}, {Brown},
  {Cantrall}, {Conrad}, {Couture}, {Delaney}, {Epps}, {Hilyard}, {Hilyard},
  {Horn}, {Jern}, {Kanto}, {Keane}, {Kibrick}, {Lewis}, {Osborne},
  {Pardeilhan}, {Pfister}, {Ricketts}, {Robinson}, {Stover}, {Tucker}, {Ward},
  \& {Wei}}]{1994SPIE.2198..362V}
{Vogt}, S.~S., {Allen}, S.~L., {Bigelow}, B.~C., {et~al.} 1994, in Society of
  Photo-Optical Instrumentation Engineers (SPIE) Conference Series, Vol. 2198,
  Instrumentation in Astronomy VIII, ed. D.~L. {Crawford} \& E.~R. {Craine},
  362, \dodoi{10.1117/12.176725}

\bibitem[{{Wagg} {et~al.}(2022{\natexlab{a}}){Wagg}, {Breivik}, \& {de
  Mink}}]{LEGWORK_joss}
{Wagg}, T., {Breivik}, K., \& {de Mink}, S. 2022{\natexlab{a}}, The Journal of
  Open Source Software, 7, 3998, \dodoi{10.21105/joss.03998}

\bibitem[{{Wagg} {et~al.}(2022{\natexlab{b}}){Wagg}, {Breivik}, \& {de
  Mink}}]{LEGWORK_apjs}
{Wagg}, T., {Breivik}, K., \& {de Mink}, S.~E. 2022{\natexlab{b}}, \apjs, 260,
  52, \dodoi{10.3847/1538-4365/ac5c52}

\bibitem[{{Yamaguchi} {et~al.}(2023){Yamaguchi}, {El-Badry}, {Fuller},
  {Latham}, {Cargile}, {Mazeh}, {Shahaf}, {Bieryla}, {Buchhave}, \&
  {Hobson}}]{Yamaguchi2023}
{Yamaguchi}, N., {El-Badry}, K., {Fuller}, J., {et~al.} 2023, arXiv e-prints,
  arXiv:2309.15905, \dodoi{10.48550/arXiv.2309.15905}

\end{thebibliography}
\bibliographystyle{aasjournal}



\end{document}